\def\Roman#1{\uppercase\expandafter{\romannumeral#1}}
\documentclass[a4paper,12pt]{article}
\usepackage{cite}
\usepackage{amssymb,amsfonts}
\usepackage{amsmath}
\usepackage[mathscr]{eucal}

\usepackage{mathrsfs}
\usepackage[utf8x]{inputenc}

\title{On the geometric representation of the  path integral reduction Jacobian for a mechanical system with symmetry given on a manifold that is a product of the total space of the principal fiber  bundle and the vector space}

\author{S. N. Storchak\\
\small{ A. A. Logunov Institute for High Energy Physics}\\
\small{of NRC ``Kurchatov Institute'',}\\
\small{Protvino, 142281, Russian Federation,}\\
}

\begin{document}

  \maketitle

\begin{abstract}
For the Jacobian resulting  from the previously considered problem of the path integral reduction in Wiener path integrals for a mechanical system with symmetry describing the motion of two interacting scalar particles on a manifold that is the product of a smooth compact finite-dimensional Riemannian manifold and a finite-dimensional vector space, a geometric representation is obtained.
This representation follows from the formula  for the scalar curvature of the original manifold endowed by definition with a free isometric smooth action of a compact semisimple Lie group.
The derivation of this formula is performed using adapted coordinates, which can be determined in the principal fiber  bundle associated with the problem under the study. 
These coordinates are similar to those used in the standard approach to quantization of Yang-Mills fields interacting with scalar fields.
\begin{flushleft}{\bf{KeyWords:}} Marsden-Weinstein reduction, Kaluza-Klein theories, Path integral, Stochastic analysis.\\
{\bf{MSC:}} 81S40 53B21 58J65
\end{flushleft}
\end{abstract}
\section{Introduction}

It is known that some finite-dimensional mechanical systems with symmetry resemble dynamical systems with gauge degrees of freedom. An example of such a system is a dynamical system that describes the motion of a scalar particle on a compact Riemannian manifold with a given free isometric action of a compact semisimple Lie group. As in gauge theories, the configuration space of this system is    the total space of the principal fiber  bundle. 
The reduction of the system leads to a mechanical system defined on the orbit space of this bundle.  The same is true for gauge theories. Therefore, one can hope that this mechanical system can be useful in studying methods for quantizing gauge fields, especially methods based on the use of path integrals.

The reduction procedure in the  Wiener path integrals\footnote{These path integrals are used for Euclidean quantization.}
 for the mechanical system under consideration, which is a model system for the pure Yang-Mills theory, was considered in \cite{Storchak_1,Storchak_11,Storchak_12,Storchak_2}. In these articles, path integrals were determined by the method of the Belopolskaya and Daletskij \cite{Dalecky_1,Dalecky_2}, with which  measures of these path integrals are generated by the stochastic processes defined on the manifold.
The stochasic processes of this definition are  solutions of the local stochastic differential equations.  These local stochastic processes are used to determine the local evolution semigroups acting in the corresponding space of functions.  In turn, the local evolutionary semigroup has a representation through the path integral whose path integral measure is given by the probability distribution of the stochastic process. The path integral on a manifold (the global path integral) is determined by the global evolution semigroup, which is the limit of the superposition of local evolution semigroups on the manifold

By the reduction procedure the original path integral was transformed to the path integral for the mechanical system whose configuration space  is the orbit space of the principal fiber bundle. This was done by factorizing the measure in the path integral. For this, a nonlinear filtering equaobtained tion from the theory of stochastic processes was applied \cite{Lipcer,Pugachev}.

The non-invariance of the path integral measure under reduction  leads to the additional term  to the potential of the reduced dynamical system. This term is the Jacobian of the reduction procedure performed in the path integral.  A geometric representation of this Jacobian was found later in \cite{Storchak_7, Storchak_8}.

The path integral reduction procedure for a mechanical system, which can be regarded as a model system for a gauge system that describes the interaction of the scalar field with a Yang-Mills field, was considered in \cite{Storchak_9}. 
The model system of this case consists of two interacting scalar particles that move on a  manifold, which is the product of a smooth compact finite-dimensional Riemannian manifold and a finite-dimensional vector space.  As in the previous case, this manifold is endowed with a free isometric smooth action of a compact semisimple Lie group. It is also assumed that the model system is invariant  under the action of this group. 

It was shown in \cite{Storchak_9} that the path integral measure is not invariant under the reduction procedure and, therefore, the potential of the reduced Hamiltonian operator has a quantum correction -- the reduction Jacobian. The purpose of our study in this article is to identify the geometric structure of this Jacobian.

\section{Definitions}
 The path integrals we were dealing with in \cite{Storchak_9} are used to represent solutions of the
 backward Kolmogorov equation given on a smooth compact Riemannian manifold $ \tilde {\mathcal P} = \mathcal P \times \mathcal V $:
\begin{equation}
\left\{
\begin{array}{l}
\displaystyle
\left(
\frac \partial {\partial t_a}+\frac 12\mu ^2\kappa \bigl[\triangle
_{\cal P}(p_a)+\triangle
_{\cal V}(v_a)\bigr]+\frac
1{\mu ^2\kappa m}V(p_a,v_a)\right){\psi}_{t_b} (p_a,t_a)=0,\\
{\psi}_{t_b} (p_b,v_b,t_b)=\phi _0(p_b,v_b),
\qquad\qquad\qquad\qquad\qquad (t_{b}>t_{a}),
\end{array}\right.
\label{1}
\end{equation}
where $\mu ^2=\frac \hbar m$ , $\kappa $  is a real positive
parameter,  $V(p,f)$ is the group-invariant potential term:
 $V(pg,g^{-1}v)=V(p,v)$, $g\in \mathcal G$, 
$\triangle _{\cal P}(p_a)$ and  $\triangle
_{\cal V}(v)$ are the Laplace--Beltrami operators on a
manifold $\cal P$ and  the vector space $\cal V$. 
In coordinates $(Q^A,f^a)$ of the point $(p,v)$ belonging to the local chart $(U_{\cal P}\times U_{\cal V},\varphi )$ of the manifold $\tilde{\mathcal P}$,
$\triangle _{\cal P}$ has the following form\footnote{In  our formulas  we  assume that there is 
sum over the repeated indices. The indices denoted by the capital
letters  run from 1 to $n_{\cal P}=\rm{dim} \cal P$, and the small Latin letters, except $i,j,k,l$, -- from 1 to $n^{\cal V}=\dim \cal V$.}:
\begin{equation}
\triangle _{\cal P}(Q)=G^{-1/2}(Q)\frac \partial {\partial
Q^A}G^{AB}(Q)
G^{1/2}(Q)\frac\partial {\partial Q^B},
\label{2}
\end{equation}
where $G=det (G_{AB})$, $G_{AB}(Q)$ is the initial Riemannian metric on $\cal P$   in the coordinate basis $\{\frac{\partial}{\partial
Q^A}\}$.
The operator $\triangle_{\cal V}$ is given by 
\[
 \triangle_{\cal V}(f)=G^{ab}\frac{\partial}{\partial f^a\partial f^b}.
\]
It is assumed that  matrix $G_{ab}$ representing the metric on $\mathcal V$
 consists of fixed constant elements. In addition, it is also assumed that $ G_{ab}$ may have off-diagonal elements.

If the necessary smooth requirements imposed on the terms of the equation are satisfied, then,  according to
\cite{Dalecky_1},  the solution of  equation  (\ref{1}) , can be represented as follows:
\begin{eqnarray}
{\psi}_{t_b} (p_a,v_a,t_a)&=&{\rm E}\Bigl[\phi _0(\eta_1 (t_b),\eta_2(t_b))\exp \{\frac
1{\mu
^2\kappa m}\int_{t_a}^{t_b}V(\eta_1(u),\eta_2(u))du\}\Bigr]\nonumber\\
&=&\int_{\Omega _{-}}d\mu ^\eta (\omega )\phi _0(\eta (t_b))\exp
\{\ldots 
\},
\label{orig_path_int}
\end{eqnarray}
where ${\eta}(t)=(\eta_1(t),\eta_2(t))$ is a global stochastic process on a manifold 
$\tilde{\cal P}=\cal P\times \cal V$; ${\mu}^{\eta}$ is the path integral measure  on
the path space $\Omega _{-}=\{\omega (t)=\omega^1(t)\times\omega^2(t):\omega^{1,2} (t_a)=0, \eta_1
(t)=p_a+\omega^1 (t),\eta_2(t)=v_a+\omega^2(t)\}$ given on manifold $\tilde {\mathcal P}$.

In a local chart  of the manifold $\tilde {\mathcal P}$, the process $\eta (t)$ is given by the solution of two stochasic differential equations:
\begin{equation}
d\eta_1^A(t)=\frac12\mu ^2\kappa G^{-1/2}\frac \partial {\partial
Q^B}(G^{1/2}G^{AB})dt+\mu \sqrt{\kappa }{\cal X}_{\bar{M}}^A(\eta_1
(t))dw^{
\bar{M}}(t),\\
\label{eta_1}
\end{equation}
and
\begin{equation}
 d\eta_2^a(t)=\mu \sqrt{\kappa }{\cal X}_{\bar{a}}^b 
dw^{
\bar{b}}(t).\\
\label{eta_2}
\end{equation}
${\cal X}_{\bar{M}}^A$ and ${\cal X}_{\bar{a}}^b$ are  defined  by the local equalities
$\sum^{n_{\mathcal P}}_{\bar{\scriptscriptstyle K}\scriptscriptstyle =1}{\cal
X}_{\bar{K}}^A{\cal X}_{\bar{K}}^B=G^{AB}$ and  $\sum^{n_{\mathcal V}}_{\bar{\scriptscriptstyle a}\scriptscriptstyle =1}{\cal
X}_{\bar{a}}^b{\cal X}_{\bar{a}}^c=G^{bc}$, $dw^{\bar{M}}(t)$ and $dw^{\bar{b}}(t)$ are the independent Wiener processes.
Here   we  denote the Euclidean 
indices by over-barred indices.

The geometry of the problem under consideration is determined by the principal fiber bundle   $ \pi':\mathcal P\times \mathcal V \to \mathcal P\times _{\mathcal G}\mathcal V$   \cite{AbrMarsd,Storchak_4,Storchak_5,Storchak_6}. It follows that the initial manifold $\tilde{\mathcal P}$ can be  viewed as  a total space of this principal fiber bundle, which is denoted by  $\rm P(\tilde{\mathcal M},\mathcal G)$.  The orbit space manifold $\tilde{\mathcal M}=\mathcal P\times _{\mathcal G}\mathcal V$  is  the base space of  $ \pi'$.
Therefore, it is possible to replace the local coordinates $(Q^A, f^n)$ of the point $ (p, v) \in \tilde{\mathcal P}$ with the coordinates given on the principal fiber bundle. As the coordinates in the  bundle space the adapted coordinates were used \cite{Creutz,Razumov_1,Razumov_3,Huffel-Kelnhofer,Kelnhofer_2}. 

These coordinates were determined using the local section $\Sigma$ chosen in the total space of the principal fiber bundle $\pi:\mathcal P\to \mathcal M $,  $\mathcal M= \mathcal P/\mathcal G$.       $\Sigma$, the local submanifold in $\mathcal P$, is given by the system of equations $\chi^{\alpha}(Q)=0,\,\alpha=1,...,n^{\mathcal G}$. 
Moreover, it was assumed that $\Sigma$ can be defined  parametrically: $Q^A=Q^{\ast A}(x^i)$, where  $x^i$, $i=1,...,n^{\mathcal M}$ are the invariant coordinates,  which can be identified with the coordinates given on the base manifold $\mathcal M$.

 The point $(p,v)\in \tilde{\mathcal P}$, which has the  coordinates e $(Q^A,f^b)$, obtains the adapted coordinates  $(x^i(Q),\tilde f^a(f), a^{\alpha}(Q))$ \cite{Storchak_6}. 
 The group coordinates $a^{\alpha}(Q)$ of a point $p\in \mathcal P$ are defined by the solution of the following equation:
\[
 \chi^{\alpha}(F^A(Q,a^{-1}(Q)))=0.
\]
(The group element carries the point $p$ to the submanifold  $\Sigma$.\footnote{$F^A(Q,g)$ denotes the right action of the group $\mathcal G$ on ${\mathcal P}$. On   the vector space $\mathcal V$, such an action  is defined as follows: ${\tilde f}^b=\bar D^b_a(g)f^a$, where $\bar D^b_a(g)\equiv D^b_a(g^{-1})$ and 
 $D^b_a(g)$ is the  matrix of  the finite-dimensional representation of the group $\mathcal G$.} )

 Invariant coordinates  $x^i(Q)$ can be determined from the equation
 \[
 Q^{\ast A}(x^i)=F^A(Q,a^{-1}(Q))),
 \]
 and $\tilde{f^b}(Q)$ are given by
\[
\tilde{f^b}(Q) = D^b_c(a(Q))\,f^c.
\]
The original metric of the Riemannian manifold $\tilde{\mathcal P}$ in the coordinate basis
$\{\partial/\partial x^i, \partial/\partial \tilde f^b, \partial/\partial a^{\alpha}\}$ has the following form:
\begin{equation}
\displaystyle
G_{\tilde A \tilde B}=
\left(
\begin{array}{ccc}
{\tilde h}_{ij}+{\mathscr A}_i^\mu {\mathscr A}_j^\nu d_{\mu \nu } & 0 & {\mathscr A}_i^\mu
d_{\mu\nu}
\bar{u}^{\nu}_{\alpha}(a) \\ 
0 & G_{ab}  & {\mathscr A}^{\mu}_a d_{\mu\nu}{\bar u}^{\nu}_{\alpha}(a)\\
{\mathscr A}^{\mu}_jd_{\mu\nu}{\bar u}^{\nu}_{\beta}(a) & {\mathscr A}^{\mu}_b d_{\mu\nu}{\bar u}^{\nu}_{\beta}(a) & d_{\mu \nu}{\bar u}^{\mu}_{\alpha}(a){\bar u}^{\nu}_{\beta}(a)\\
\end{array}
\right),
\label{transfmetric}
\end{equation}
where 
${\tilde h}_{ij}(x,\tilde f)=Q^{\ast}{}^A_i{\tilde G}^{\rm H}_{AB}Q^{\ast}{}^B_j$ with ${\tilde G}^{\rm H}_{AB}=G_{AB}-G_{AC}K^C_{\mu}d^{\mu\nu}K^D_{\nu}G_{DB}$,  ${\tilde G}^{\rm H}_{AB}=\tilde{\Pi}^C_AG_{CB}$. $({\mathscr A}^{\alpha}_i, {\mathscr A}^{\alpha}_p)$ are   components of the mechanical connection 
$${\omega}^{\alpha}=\bar\rho ^{\alpha}_{\beta}(a)({{\mathscr A}}^{\beta}_idx^i+{{\mathscr A}}^{\beta}_cd\tilde f^c) +u^{\alpha}_{\nu}da^{\nu}$$
 existing in $\rm P(\tilde{\mathcal M},\mathcal G)$. They are determined as 
  $${\mathscr A}^{\alpha}_i(x,\tilde f)=d^{\alpha \beta}K^C_{\beta}G_{DC}Q^{\star}{}^D_i, \;\;\;{\mathscr A}^{\alpha}_c(x,\tilde f)=d^{\alpha \beta}K^a_{\beta}G_{ac}.$$

$K^A_{\beta}$ and $K^a_{\beta}$ are the Killing vector fields for the metric given on $\tilde{\mathcal P}$.  $d_{\mu\nu}$ is  metric on the orbits in  $\rm P(\tilde{\mathcal M},\mathcal G)$:
$$d_{\mu \nu}(x,\tilde f)={\gamma}_{\mu\nu}(x)+{\gamma'}_{\mu\nu}(\tilde f)=G_{AB}K^A_{\mu}K^B_{\nu}+G_{ab}K^a_{\mu}K^b_{\nu}.$$
And $\bar{\rho}^{\alpha}_{\beta}$ is an inverse matrix to the matrix ${\rho}^{\alpha}_{\beta}=\bar{u}^{\alpha}_{\sigma}{v}^{\sigma}_{\beta}$ of an adjoint representation of the group $\mathcal G$.

Further, we will also denote expressions that include $d^{\mu\nu}$, with a tilde mark above the character associated with that expression.

The inverse matrix $G^{\tilde A \tilde B}$ to matrix (\ref{transfmetric}) is as follows:
\begin{equation}
 \displaystyle
G^{\tilde A \tilde B}=\left(
\begin{array}{ccc}
 h^{ij} & \underset{\scriptscriptstyle{(\gamma)}}{{\mathscr A}^{\mu}_m} K^a_{\mu} h^{mj} & -h^{nj}\,\underset{\scriptscriptstyle{(\gamma)}}{{\mathscr A}^{\beta}_n} \bar v ^{\alpha}_{\beta} \\
\underset{\scriptscriptstyle{(\gamma)}}{{\mathscr A}^{\mu}_m} K^b_{\mu} h^{ni} & G^{AB}N^a_AN^b_B+G^{ab} & -G^{EC}{\Lambda}^{\beta}_E{\Lambda}^{\mu}_CK^b_{\mu}\bar v ^{\alpha}_{\beta}
\\
-h^{ki}\underset{\scriptscriptstyle{(\gamma)}}{{\mathscr A}^{\varepsilon}_k}\bar v ^{\beta}_{\varepsilon} & -G^{EC}{\Lambda}^{\varepsilon}_E{\Lambda}^{\mu}_CK^a_{\mu}\bar v ^{\beta}_{\varepsilon} & G^{BC}{\Lambda}^{\alpha'}_B{\Lambda}^{\beta'}_C\bar v ^{\alpha}_{\alpha'}v ^{\beta}_{\beta'}
\end{array}
\right),
\label{invers_metric}
\end{equation}
where ${\Lambda}^{\beta}_E=(\Phi^{-1})^{\beta}_{\mu}{\chi}^{\mu}_E$,  $h^{ij}$ is an inverse matrix to the matrix $h_{ij}=Q^{\ast A}_i G^{\rm H}_{AB}Q^{\ast B}_j$ with $$G^{\rm H}_{AB}=G_{AB}-G_{AD}K^D_{\alpha}{\gamma}^{\alpha\beta}K^C_{\beta}G_{CB}.$$ 
Also, for the mechanical connection in  $\rm P({\mathcal M},\mathcal G)$, where the orbit metric  is $\gamma_{\mu \nu}$, we use the following notation:
$$\underset{\scriptscriptstyle{(\gamma)}}{{\mathscr A}^{\mu}_m}={\gamma}^{\mu \nu}K^A_{\nu}G_{AB}Q^{\ast B}_m.$$
By $\chi^{\alpha}_B$ we denote $\chi^{\alpha}_B=\partial \chi^{\alpha}(Q)/\partial Q^B |_{Q=Q^{\ast}(x)}$, $(\Phi)^{\alpha}_{\beta}=K^A_{\beta}\chi^{\alpha}_A$ is the Faddeev-Popov matrix, $N^b_B=-K^b_{\mu}(\Phi)^{\mu}_{\nu}{\chi}^{\nu}_B \equiv -K^b_{\mu}{\Lambda}^{\mu}_B$ is one of the components of a particular projector on a tangent space to the orbit space $\tilde{\mathcal M}$. This projector $N=(N^A_B=\delta^A_B-K^A_{\mu}{\Lambda}^{\mu}_B, N^b_B, N^A_b=0,N^a_b=\delta^a_b)$ was defined in \cite{Storchak_4,Storchak_5}.

The determinant of the matrix (\ref{transfmetric}) consists of three multipliers:
\begin{eqnarray}
 \det G_{\tilde A \tilde B}=(\det d_{\alpha\beta})\,(\det {\bar u}^{\mu}_{\nu}(a))^2 \displaystyle
\det \left(
\begin{array}{cc}
\tilde h_{ij} & \tilde G^{\rm H}_{B b}Q_{i}^{*B}\\
\tilde G^{\rm H}_{Aa}Q_{j}^{*A} & \tilde G^{\rm H}_{ba}\\
\end{array}
\right),
\label{det}
\end{eqnarray}
where $\tilde G^{\rm H}_{Aa}=-G_{AB}K^B_{\mu}d^{\mu\nu}K^b_{\nu}G_{ba}$, $\tilde G^{\rm H}_{ba}=G_{ba}-G_{bc}K^c_{\mu}d^{\mu\nu}K_{\nu}^pG_{pa}$.

The last determinant on the right hand side of (\ref{det}) is the determinant of the metric  defined on the orbit space $\tilde{\mathcal M}=
\mathcal P\times_{\mathcal G}\mathcal V$ of the principal fiber bundle $\rm P(\tilde{\mathcal M},\mathcal G).$
 In the paper, this determinant is denoted  by $H(x,\tilde f)$.

Note that the upper left quadrant of the matrix
(\ref{invers_metric}) is a matrix that represents the inverse metric to the metric on the orbit space 
 of our principal fiber bundle.

As a result of the path integral transformation, which consists of the transformation of the stochastic process (transition to the adapted variables) and making use of the nonlinear filtering stochastic differential equation followed by the Girsanov transformation, we obtained\cite{Storchak_9}  the integral relation between the path integral given on $\tilde{\mathcal M}$ and the path integral defined on the total space of the principal fiber bundle $\tilde{\mathcal P}$. For the zero-momentum level reduction this relation is as follows:
\begin{equation}
d_b^{-1/4}d_a^{-1/4}
G_{\tilde{\scriptscriptstyle\mathcal M}}(x_b,\tilde f_b, t_b;x_a,\tilde f_a,t_a)=\int_{\mathcal G}{G}_{\tilde{\scriptscriptstyle\mathcal P}}(p_b\theta,v_b\theta,t_b;p_a,v_a,t_a)
d\mu (\theta ),
\nonumber\\
\end{equation}
$(x,\tilde f)=\pi'(p,v)$, and $d_b=d(x_b,\tilde f_b)$, $d_a=d(x_a,\tilde f_a)$. $d\mu (\theta) $ is a normalized invariant Haar measure on a group $\mathcal G$, ($\int_{\mathcal G}d\mu (\theta )=1$).
 
 The Green function ${G}_{\tilde{\scriptscriptstyle\mathcal P}}(Q_b, f_b,t_b;Q_a,f_a,t_a)$ representing the kernel of the evolution semigroup (\ref{orig_path_int}) acts in the Hilbert space of functions with the scalar product $(\psi_1,\psi_2)=\int \psi_1(Q,f)\psi_2(Q,f)dv_{\tilde{\scriptscriptstyle\mathcal P}}(Q,f),$ ($v_{\tilde{\mathcal P}}$ is a volume measure on $\tilde{\mathcal P}$, $dv_{\tilde{\mathcal P}}(Q,f)=\sqrt{G(Q,f)}dQ^1\dots dQ^{n_{\mathcal P}}df^1\dots df^{n_{\mathcal V}}$).
  
The semigroup which is determined by the Geen function $G_{\tilde{\scriptscriptstyle\mathcal M}}$ acts in the Hilbert space with the scalar product $(\psi_1,\psi_2)=\int \psi_1(x,\tilde f),\psi_2(x,\tilde f)dv_{\tilde{\mathcal M}}$, where $dv_{\tilde{\mathcal M}}=\sqrt{H(x,\tilde f)}dx^1\dots dx^{n_{\scriptscriptstyle\mathcal M}}d\tilde f^1\dots d\tilde f^{n_{\mathcal V}}$.

The Green function $G_{\tilde{\scriptscriptstyle\mathcal M}}$ is given by the following path integral:
\begin{eqnarray}
&&G_{\tilde{\scriptscriptstyle\mathcal M}}(x_b,\tilde f_b, t_b;x_a,\tilde f_a,t_a)
\nonumber\\
&&\;\;\;\;\;\;\;\;\;\;\;\;\;=\int_{ 
{\tilde{\xi}(t_a)=(x_a,\tilde f_a)}\atop
{\tilde{\xi}(t_b)=(x_b,\tilde f_b)}}
d\mu ^{\tilde{\xi}}\exp 
\left\{\frac 1{\mu
^2\kappa m}\int_{t_a}^{t_b}
\tilde V(\tilde{\xi}_1(u),\tilde{\xi}_2(u))du\right\}
\nonumber\\
&&\;\;\;\;\;\;\;\;\;\;\;\;\;\times\exp\Bigl\{-\frac18\mu^2\kappa\int^{t_b}_{t_a}\bigl(\triangle_{\tilde{\scriptscriptstyle\cal M}}\sigma +\frac14<\partial\sigma,\partial \sigma>_{\tilde{\scriptscriptstyle\cal M}}\bigr)du\Bigr\},
\nonumber\\
&&\;\;\;\;\;\;\;\;\;\;\;\;\;(x,\tilde f)=\pi'(p,v),  \;\;\sigma =\sigma(\tilde{\xi}_1(u),\tilde{\xi}_2(u)), 
\label{path_int_M}
\end{eqnarray}
where $\tilde V(x,\tilde f)=V(F(Q^{\ast}(x),a),\bar D(a)\tilde f)$
  and 
 $<\partial\sigma,\partial \sigma>_{\tilde{\scriptscriptstyle\cal M}}$ is the quadratic form  obtained using the  metric  on $\tilde{\cal M}$:
\[
 \Bigl[h^{ij}\sigma_i\sigma_j+2h^{kj}\underset{\scriptscriptstyle{(\gamma)}}{{\mathscr A}^{\mu}_k} K^a_{\mu} \sigma_a\sigma_j+\Bigl((\gamma^{\alpha\beta}+h^{kl}\underset{\scriptscriptstyle{(\gamma)}}{{\mathscr A}^{\alpha}_k}  \underset{\scriptscriptstyle{(\gamma)}}{{\mathscr A}^{\beta}_l}) K^a_{\alpha}K^b_{\beta}+G^{ab}\Bigr)\sigma_a\sigma_b\Bigr],
\]
 $\sigma_i=\frac{\partial}{\partial x^i} (\ln d)$ and $\sigma_a=\frac{\partial}{\partial \tilde f^a} (\ln d)$. 

The measure in the path integral (\ref{path_int_M}) is generated by the stochastic process $\tilde{\xi}=( \tilde{\xi}_1,\tilde{\xi}_2)$   on the manifold $\tilde{\mathcal M}$.  The local stochasic differential equation of this process is
\[
 d\tilde\xi_{\rm loc}(t)=\frac12\mu^2\kappa{\tilde b^i\choose \tilde b^a}dt+\mu\sqrt{\kappa}{\tilde X^i_{\bar m}\;\; 0\choose {\tilde X^a_{\bar m}\;\; \tilde X^a_{\bar b}}} {d\tilde w^{\bar m}\choose d\tilde w^{\bar b}}.
\]
The drift coefficients of this equation are given by the following expressions:
\[
 \tilde b^i=\frac1{\sqrt{H}}\frac{\partial}{\partial x^j}\Bigl(\sqrt{H}h^{ij}\Bigr)+\underset{\scriptscriptstyle{(\gamma)}}{{\mathscr A}^{\mu}_n}h^{ni}\frac1{\sqrt{H}}\frac{\partial}{\partial \tilde f^b}\Bigl(\sqrt{H}K^b_{\mu}\Bigr)
\]
and
\begin{eqnarray*}
 \tilde b^a&=&\frac1{\sqrt{H}}\frac{\partial}{\partial x^j}\Bigl(\sqrt{H}h^{mj}\underset{\scriptscriptstyle{(\gamma)}}{{\mathscr A}^{\mu}_m}\Bigr)K^a_{\mu}+(G^{ab}+G^{AB}N^a_AN^b_B)\frac1{\sqrt{H}}\frac{\partial}{\partial \tilde f^b}\Bigl(\sqrt{H}\Bigr)
\\
&&+\frac{\partial}{\partial \tilde f^b}\Bigl(G^{AB}N^a_AN^b_B\Bigr).
\end{eqnarray*}
The diffusion  coefficients are as folows:
$${\tilde X}^i_{\bar m}=(h^{ij})^{1/2},\;\;\; {\tilde X}^{a}_{\bar m}={\tilde X}^k_{\bar m}\underset{\scriptscriptstyle{(\gamma)}}{{\mathscr A}^{\mu}_k} K^a_{\mu},
 \;\;\;\tilde X^a_{\bar b}=(\gamma^{\alpha \beta}K^a_{\alpha}K^b_{\beta}+G^{ab})^{1/2}.$$
Note that with respect to the variables $ (x_b, \tilde f_b, t_b) $, the Green function $ G_{\tilde{\mathcal M}} $, which is the kernel of the reduced evolution semigroup,  satisfies the forward Kolmogorov equation. The differential operator $\hat{H}_{\kappa}$ of this equation (differential generator of a reduced semigroup) has the following form:
\[
\hat{H}_{\kappa}=
\frac{\hbar \kappa}{2m}\triangle _{\scriptscriptstyle\tilde{\mathcal M}}-\frac{\hbar \kappa}{8m}\Bigl[\triangle_{\scriptscriptstyle\tilde{\cal M}}\sigma +\frac14<\partial\sigma,\partial \sigma>_{\tilde{\scriptscriptstyle\cal M}}\Bigr]+\frac{1}{\hbar \kappa}\tilde V,
\]
where $\triangle _{\scriptscriptstyle\tilde{\mathcal M}}$ is the Laplace-Beltrami operator on the manifold $\tilde{\mathcal M}$.
Assuming k = i, the forward Kolmogorov equation can be rewritten as the Schr\"odinger equation with the Hamilton operator
$\hat{H}_{\tilde{\scriptscriptstyle\mathcal M}}=-\frac{\hbar}{\kappa}{\hat H}_{\kappa}\bigl|_{\kappa =i}$.

Thus, the measure in the  path integral (\ref{orig_path_int}) is not invariant under the  reduction procedure.
and the reduction  Jacobian is
\begin{equation}
J=-\frac{\mu^2\kappa}{8}\Bigl[\triangle_{\scriptscriptstyle\tilde{\cal M}}\sigma +\frac14<\partial\sigma,\partial \sigma>_{\tilde{\scriptscriptstyle\cal M}}\Bigr]\equiv-\frac{\mu^2\kappa}{8}\tilde J(x,\tilde f).
\label{jacobian}
\end{equation}
 This is  the result of the calculations performed  in \cite{Storchak_9}.
 In the next section,  we proceed to the search for a geometrical representation of this Jacobian.

\section{Jacobian}
		In this section, we will show that the geometrical  representation of the  Jacobian (\ref{jacobian}) follows from the formula for the scalar curvature of the Riemannian manifold $ \tilde{\mathcal P} $ with the Kaluza-Klein metric (\ref{transfmetric}). The formula is obtained in Appendix B using the Christoffel coefficients, which were calculated in Appendix A. This was done using a special basis, a horizontal lift basis $(\hat H_i,\hat H_a,L_{\alpha})$, in which the metric (\ref{transfmetric}) is written as a block diagonal matrix. As a result, we get the following representation for the scalar curvature ${\rm R}_{\tilde{\mathcal P}}$:

\begin{eqnarray}
&&\!\!\!{\rm R}_{\tilde{\mathcal P}}={\rm R}_{\tilde{\mathcal M}}+{\rm R}_{\mathcal G}+\frac14 \tilde h^{A'B'}\tilde h^{C'D'}d_{\mu\nu}{\mathscr F}^{\mu}_{A'C'}{\mathscr F}^{\nu}_{B'D'}
\nonumber\\
&&\!\!\!+\frac14\tilde h^{A'B'}d^{\mu\sigma}d^{\nu\kappa}({\mathscr D}_{A'}d_{\mu\nu})({\mathscr D}_{B'}d_{\sigma\kappa})+\triangle_{\scriptscriptstyle \tilde {\mathcal M}}\ln d+\frac14G_{\scriptscriptstyle \tilde{\mathcal M}}(\partial \ln d,\partial \ln d).
\label{curvat_itog}
\end{eqnarray}
(The capital letters with a prime as a superscript that we have used here in indices, mean the condensed notation, for example, $ A '= (i, a) $, etc.)

In the obtained formula,   
$R_{\mathrm {\cal G}}=\frac12{d}^{\mu\nu} c^{\sigma}_{\mu \alpha} c^{\alpha}_{\nu\sigma}+
\frac14 {d}_{\mu\sigma}{d}^{\alpha\beta}{d}^{\epsilon\nu}
c^{\mu}_{\epsilon \alpha}c^{\sigma}_{\nu \beta}$ is the scalar curvature of the orbit, the covariant derivative ${\mathscr D}_{A'}d_{\mu\nu}$ is given by
${\mathscr D}_{\scriptscriptstyle A'}d_{\mu\nu}=\partial_{\scriptscriptstyle A'}d_{\mu\nu}-c^{\kappa}_{\sigma \mu}\mathscr A^{\sigma}_{\scriptscriptstyle A'}d_{\kappa\nu}-c^{\kappa}_{\sigma \nu}\mathscr A^{\sigma}_{\scriptscriptstyle A'}d_{\mu\kappa}$ and the curvature ${\mathscr F}^{\mu}_{\scriptscriptstyle A'C'}$ is defined as ${\mathscr F}^{\mu}_{\scriptscriptstyle A'C'}=\partial_{\scriptscriptstyle A'}\mathscr A^{\mu}_{\scriptscriptstyle C'}-\partial_{\scriptscriptstyle C'}\mathscr A^{\mu}_{\scriptscriptstyle A'}+c^{\mu}_{\sigma\nu}\mathscr A^{\sigma}_{\scriptscriptstyle A'}\mathscr A^{\nu}_{\scriptscriptstyle C'}$. The Laplace-Beltrami operator   is $ \triangle_{\scriptscriptstyle \tilde {\mathcal M}}=\tilde h^{\scriptscriptstyle A'B'}\partial_{\scriptscriptstyle A'}\partial_{\scriptscriptstyle B'}-\tilde h^{\scriptscriptstyle A'B'}\Gamma^{\scriptscriptstyle C'}_{\scriptscriptstyle A'B'}\partial_{\scriptscriptstyle C'}$, $\tilde h^{\scriptscriptstyle A'B'}$ are the elements of the upper left quadrant of the matrix (\ref{invers_metric}).

From the obtained expression for the curvature ${\rm R}_{\tilde{\mathcal P}}$ we see that the two last terms are equal  to expression $\tilde J$ of the Jacobian  (\ref{jacobian}). So, we have
\begin{equation}
\tilde J={\rm R}_{\tilde{\mathcal P}}-{\rm R}_{\tilde{\mathcal M}}-{\rm R}_{\mathcal G}-\frac14 d_{\mu\nu}{\mathscr F}^{\mu}_{\scriptscriptstyle A'\scriptscriptstyle B'}{\mathscr F}^{\nu \scriptscriptstyle A'\scriptscriptstyle B'}-\frac14\tilde h^{\scriptscriptstyle A'\scriptscriptstyle B'}d^{\mu\sigma}d^{\nu\kappa}({\mathscr D}_{\scriptscriptstyle A'}d_{\mu\nu})({\mathscr D}_{\scriptscriptstyle B'}d_{\sigma\kappa}).
\label{jacobian_1}
\end{equation}
The last expression on the right-hand side (\ref{jacobian_1}) also has a geometric representation, which, as will be shown below, is associated with the second fundamental form of the orbit in the principal fiber bundle $\rm P(\tilde{\mathcal M},\tilde G)$.

In the total space of the bundle this form is determined as 
\begin{eqnarray}
j_{\alpha\beta}(Q,f)&=&\Bigl(\tilde \Pi^D_C({\nabla}_{K_{\alpha}}K_{\beta})^C(Q)+{\tilde \Pi}^D_b({\nabla}_{K_{\alpha}}K_{\beta})^b(f)\Bigr)\frac{\partial}{\partial Q^D}\nonumber\\
&&\!\!\!+\Bigl(\tilde \Pi^a_C({\nabla}_{K_{\alpha}}K_{\beta})^C(Q)+{\tilde \Pi}^a_b({\nabla}_{K_{\alpha}}K_{\beta})^b(f)\Bigr)\frac{\partial}{\partial f^a},
\label{j_Qf}
\end{eqnarray}
where  
$\tilde{\Pi}=(\tilde \Pi^D_C,{\tilde \Pi}^D_b,\tilde \Pi^a_C,{\tilde \Pi}^a_b)$ is the horizontal projector, the  projector ``in the direction  orthogonal to the orbit.'' The Killing vector fields
$$K_{\alpha}(Q,f)=K^A_{\alpha}(Q)\frac{\partial}{\partial Q^A}+K^a_{\alpha}(f)\frac{\partial}{\partial f^a}.$$
are tangent to the orbit.

 $({\nabla}_{K_{\alpha}}K_{\beta})^C(Q)$ can be decomposed into symmetric and antisymmetric parts:
\[({\nabla}_{K_{\alpha}}K_{\beta})^C(Q)=\frac12\Bigl[({\nabla}_{K_{\alpha}}K_{\beta})^C+({\nabla}_{K_{\beta}}K_{\alpha})^C\Bigr]+\frac12\Bigl[({\nabla}_{K_{\alpha}}K_{\beta})^C-({\nabla}_{K_{\beta}}K_{\alpha})^C\Bigr].
\]
Note that in the consequent calculations  it will be sufficient to use only the symmetric part of this expression. This conclusion can be drawn from the following.

Since there is no torsion, the expression in the second bracket can be rewritten as $\frac12 c^{\varphi}_{\alpha\beta}K^C_{\varphi}$. For the antisymmetric part of  $({\nabla}_{K_{\alpha}}K_{\beta})^b(f)$,  a similar expression  can be obtained . Then it can be shown that the antisymmetric part of $\tilde \Pi^D_CK^C_{\varphi}+\tilde \Pi^D_bK^b_{\varphi}=0,$ vanishes  due to the identity
$\tilde \Pi^D_CK^C_{\varphi}+\tilde \Pi^D_bK^b_{\varphi}=0.$
Note that the same is true for the antisymmetric part of  $({\nabla}_{K_{\alpha}}K_{\beta})^a(f)$.
 
Next, it is necessary to project the second quadratic form $ j_{\alpha \beta}$ onto ``the direction which is parallel to the the orbit space''\footnote{Projection is performed on the plane in $\tilde{\mathcal P}$ that is
 parallel to the tangent plane taken  at the point  $(x^i,\tilde f^b)$ belonging to $\tilde{\mathcal M}$.}
 $\tilde{\mathcal M}$. This can be done as follows:
$$\tilde h^{kn}\tilde G(j_{\alpha \beta}, \hat{H}_k)\hat{H}_n+\tilde h^{kb}\tilde G(j_{\alpha \beta}, \hat{H}_k)\hat{H}_b
+\tilde h^{bk}\tilde G(j_{\alpha \beta}, \hat{H}_b)\hat{H}_k
+\tilde h^{ab}\tilde G(j_{\alpha \beta}, \hat{H}_a)\hat{H}_b,$$
where $\tilde G$ is the metric (\ref{transfmetric}), $j_{\alpha \beta}$ is an expression obtained from the right-hand side of (\ref{j_Qf}) by replacing the original coordinates $ (Q^A, f^b)$ in it with $(x^i,\tilde f^b, a^{\alpha})$.
$ \hat{H}_i=\Bigl(\frac{\partial}{\partial x^i}-\tilde {\mathscr A}^{\alpha}_iL_{\alpha}\Bigr)$ and    $\hat{H}_a=\Bigl(\frac{\partial}{\partial \tilde f^a}-\tilde {\mathscr A}^{\alpha}_aL_{\alpha}\Bigr)$, where $\tilde {\mathscr A}^{\alpha}_i=\bar{\rho}^{\alpha}_{\beta} {\mathscr A}^{\beta}_i$, are the vector filds of the horizontal lift basis.

The previous four terms representing the projection $j_{\alpha\beta}$ can be written as
\begin{eqnarray}
&&j(1)_{\alpha\beta}=\Bigl[\tilde h^{kn}\tilde G^{\rm H}_{BM}Q^{\ast B}_k(\dots)^M+\tilde h^{kn}\tilde G^{\rm H}_{Ab}Q^{\ast A}_k(\dots)^b\Bigr]\hat{H}_n,
\nonumber\\
&&j(2)_{\alpha\beta}=\tilde h^{kb}\Bigl[\Bigl( G^{\rm H}_{ML}Q^{\ast L}_mh^{mi}\tilde h_{ki}+N^a_M\tilde h_{ak}\Bigr)(\dots)^M+\tilde h_{ck}(\dots)^c\Bigr]\hat{H}_b,
\nonumber\\
&&j(3)_{\alpha\beta}=\tilde h^{bk}\Bigl[\Bigl( G^{\rm H}_{ML}Q^{\ast L}_mh^{mi}\tilde h_{ib}+N^a_M\tilde h_{ab}\Bigr)(\dots)^M+\tilde h_{cb}(\dots)^c\Bigr]\hat{H}_k,
\nonumber\\
&&j(4)_{\alpha\beta}=\tilde h^{ab}\Bigl[\tilde G^{\rm H}_{Ma}(\dots)^M+\tilde G^{\rm H}_{da}(\dots)^d\Bigr]\hat{H}_b,
\label{j_1-4}
\end{eqnarray}
\begin{eqnarray*}
\rm{where} &(\dots)^M&=\frac12\Bigl(({\nabla}_{K_{\alpha}}K_{\beta})^M(Q^{\ast}(x))+({\nabla}_{K_{\beta}}K_{\alpha})^M(Q^{\ast}(x))\Bigr),
\nonumber\\
&(\dots)^d&=\frac12\Bigl(({\nabla}_{K_{\alpha}}K_{\beta})^d(\tilde f)+({\nabla}_{K_{\beta}}K_{\alpha})^d(\tilde f)\Bigr).
\end{eqnarray*}

Appendix C shows how, using identities expressing $ (\dots)^M$ and $ (\dots)^b $ in terms of the partial derivatives $d_{\alpha \beta} $, to simplify the expressions for these  components of the projection $ j_{\alpha \beta} $.
As a result of the calculation performed in this application, we get
\begin{eqnarray*}
&&j(1)_{\alpha\beta}=-\frac12\tilde h^{kn}(\mathscr D_kd_{\alpha\beta})\hat H_n,\;\;\;
j(2)_{\alpha\beta}=-\frac12\tilde h^{kb}(\mathscr D_kd_{\alpha\beta})\hat H_b,
\nonumber\\
&&j(3)_{\alpha\beta}=-\frac12\tilde h^{bk}(\mathscr D_b d_{\alpha\beta})\hat H_k,\;\;\;
j(4)_{\alpha\beta}=-\frac12\tilde h^{ab}(\mathscr D_a d_{\alpha\beta})\hat H_b.
\end{eqnarray*}

So, the second fundamental form of the orbit is
$$j_{\alpha\beta}=-\frac12\Bigl[(\tilde h^{kn}{\mathscr D}_kd_{\alpha\beta}+\tilde h^{bn}{\mathscr D}_bd_{\alpha\beta})\,\hat H_n+(\tilde h^{kb}{\mathscr D}_kd_{\alpha\beta}+\tilde h^{ab}{\mathscr \,D}_ad_{\alpha\beta})\,\hat H_b].$$

It can be shown that the necessary Jacobian term can be represented as a trace of the square $ j_{\alpha \beta}$ taken on 
 $\tilde{\mathcal M}$: 
 \begin{eqnarray*}
G_{\scriptscriptstyle\tilde{\mathcal M}}(j_{\alpha\beta},j_{\mu\nu})d^{\alpha\beta}d^{\mu\nu}&=&\frac14d^{\alpha\beta}d^{\mu\nu}\Bigl[({\mathscr D}_kd_{\alpha\beta})({\mathscr D}_ld_{\mu\nu})\tilde h^{kl}
+({\mathscr D}_bd_{\alpha\beta})({\mathscr D}_ld_{\mu\nu})\tilde h^{bl}
\nonumber\\
&&
\;\;\;+({\mathscr D}_kd_{\alpha\beta})({\mathscr D}_d
d_{\mu\nu})\tilde h^{dk}+({\mathscr D}_ad_{\alpha\beta})({\mathscr D}_d
d_{\mu\nu})\tilde h^{ad}\Bigr],
\end{eqnarray*}
$G_{\scriptscriptstyle\tilde{\mathcal M}}$ is the metric on ${\tilde{\mathcal M}}$ in the basis 
 ${\hat H_n}|_{\scriptscriptstyle\tilde{\mathcal M}}=\frac{\partial}{\partial x^n}$ and ${\hat H_a}|_{\scriptscriptstyle\tilde{\mathcal M}}=\frac{\partial}{\partial \tilde f^a}.$

Thus, we conclude that the Hamilton operator of the Schr\"odinger equation on the reduced manifold  $\tilde{\mathcal M}$  has the following form:
\[
\hat{H}_{\scriptscriptstyle\tilde{\mathcal M}}=-
\frac{\hbar ^2}{2m}\triangle _{\scriptscriptstyle\tilde{\mathcal M}}+\frac{\hbar ^2}{8m}\Bigl[{\rm R}_{\scriptscriptstyle\tilde{\mathcal P}}-{\rm R}_{\scriptscriptstyle\tilde{\mathcal M}}-{\rm R}_{\scriptscriptstyle\mathcal G}-\frac14 d_{\mu\nu}{\mathscr F}^{\mu}_{\scriptscriptstyle A'\scriptscriptstyle B'}{\mathscr F}^{\nu \scriptscriptstyle A'\scriptscriptstyle B'}-||j||^2\Bigr]+\tilde V.
\]

\appendix
\section*{Appendix A}
\section*{Christoffel symbols for the metric on $\tilde{\mathcal P}$}
\setcounter{equation}{0}
\def\theequation{A.\arabic{equation}}

In this article, the Christoffel symbols are calculated using a nonholonomic basis on the manifold $\tilde{\mathcal P}$. This basis consists of the following vector fields $(\hat H_i,\hat H_a,L_{\alpha})$, where 
$$\hat H_i=\frac{\partial}{\partial x^i}-{\tilde{\mathscr A}}^{\alpha}_iL_{\alpha},\;\;\;
\hat H_a=\frac{\partial}{\partial \tilde f^a}-{\tilde{\mathscr A}}^{\alpha}_aL_{\alpha},\;\;\;
L_{\alpha}=v^{\mu}_{\alpha}(a)\frac{\partial}{\partial a^{\mu}},$$
with ${{\mathscr A}}^{\alpha}_a(x,\tilde f)=d^{\alpha\beta}(x,\tilde f)K^b_{\alpha}(\tilde f)G_{ba}\;\; $, 
${\tilde{\mathscr A}}^{\alpha}_i(x,\tilde f,a)=\bar\rho ^{\alpha}_{\beta}(a){{\mathscr A}}^{\beta}_i$
and\\
 ${{\mathscr A}}^{\alpha}_i(x,\tilde f)=d^{\alpha\beta}(x,\tilde f)K^A_{\alpha}(Q^{\ast}(x))G_{AB}(Q^{\ast}(x))Q^{\ast B}_i(x)$.\footnote{
${\omega}^{\alpha}=\bar\rho ^{\alpha}_{\beta}({{\mathscr A}}^{\beta}_idx^i+{{\mathscr A}}^{\beta}_cd\tilde f^c) +u^{\alpha}_{\nu}da^{\nu}$.}

The commutation relations of these vector fields have the form:
$$[\hat H_i,\hat H_j]=-{\tilde{\mathscr F}}^{\gamma}_{ij}L_{\gamma},\;\;[\hat H_i,\hat H_b]=-{\tilde{\mathscr F}}^{\gamma}_{ib}L_{\gamma},\
[\hat H_a,\hat H_b]=-{\tilde{\mathscr F}}^{\gamma}_{ab}L_{\gamma},$$
$$[\hat H_i,L_{\alpha}]=0,\;\;\;[\hat H_b,L_{\alpha}]=0,\;\;\;[L_{\alpha},L_{\beta}]=c^{\gamma}_{\alpha\beta}L_{\gamma}.$$
We denote the structure constants of these commutation relation as
$$\mathbb{C}^{\gamma}_{ij}=-{\tilde{\mathscr F}}^{\gamma}_{ij},\;\;\;\mathbb{C}^{\gamma}_{ib}=-{\tilde{\mathscr F}}^{\gamma}_{ib},\;\;\;\mathbb{C}^{\gamma}_{ab}=-{\tilde{\mathscr F}}^{\gamma}_{ab},\;\;\;\mathbb{C}^{\gamma}_{\alpha\beta}=c^{\gamma}_{\alpha\beta}.$$

In the basis $(\hat H_i,\hat H_b,L_{\alpha})$, the original metric (\ref{transfmetric}) can be represented as follows:
\begin{equation}
\displaystyle
\left(
\begin{array}{ccc}
 \tilde h_{ij} & \tilde h_{ia} & 0 \\ 
  \tilde h_{bj} & \tilde h_{ba} & 0  \\
0 & 0 & {\tilde d}_{\alpha\beta}\\
\end{array}
\right),
\label{11}
\end{equation}
${\tilde d}_{\alpha\beta}=\rho ^{\alpha'}_{\alpha}\rho ^{\beta'}_{\beta}d_{\alpha'\beta'}.$

The metric on $\tilde{\mathcal M}$ is given by the matrix
\begin{equation}
\displaystyle
\left(
\begin{array}{cc}
 Q^{\ast A}_i{\tilde G}^{\rm H}_{AB}Q^{\ast B}_j & {\tilde G}^{\rm H}_{Aa}Q^{\ast A}_i  \\ 
  {\tilde G}^{\rm H}_{Bb}Q^{\ast B}_j & {\tilde G}^{\rm H}_{ba}
\end{array}
\right)=\left(
\begin{array}{cc}
 \tilde h_{ij} &  \tilde h_{ia} \\

 \tilde h_{bj} &  \tilde h_{ba} \\
\end{array}
\right),
\label{12}
\end{equation}

$ {\tilde G}^{\rm H}_{ab}=G_{ab}-G_{ac}K^c_{\alpha}d^{\alpha\beta}K^{d}_{\beta}G_{db}$.

The inverse matrix to the matrix (\ref{12}) is 
\begin{equation}
\displaystyle
\left(
\begin{array}{cc}
 {h^{ij}} & \underset{\scriptscriptstyle{(\gamma)}}{{\mathscr A}^{\mu}_m} K^a_{\mu}{h^{mj}}  \\ 
 \underset{\scriptscriptstyle{(\gamma)}}{{\mathscr A}^{\alpha'}_n}K^b_{\alpha'}{h^{ni}} & G^{ab}+G^{AB}N^a_AN^b_B\\
\end{array}
\right).
\label{13}
\end{equation}
$h^{ij}$ is an inverse to $h_{ij}=Q^{\ast}{}^A_iG^{\rm H}_{AB} Q^{\ast}{}^A_j$ with
$G^{\rm H}_{AB}=G_{AB}-G_{AC}K^C_{\alpha}{\gamma}^{\alpha\beta}K^D_{\beta}G_{DB},$
 $\underset{\scriptscriptstyle{(\gamma)}}{{\mathscr A}^{\beta}_j}$  is determined as
$ \underset{\scriptscriptstyle{(\gamma)}}{{\mathscr A}^{\beta}_j}={\gamma}^{\beta\mu}K^A_{\mu}G_{AB}Q^{\ast B}_j.$

Note  that the components of the inverse matrix (\ref{13}) can also be written as follows:
\begin{eqnarray*}
 &&\tilde h^{ji}=G^{EF}N^S_EN^D_FT^j_ST^i_D={h^{ij}}\\
&&\tilde h^{jb}=G^{EF}N^b_FN^P_ET^j_P\\
&&\tilde h^{cb}=G^{cb}+G^{EF}N^c_EN^b_F\,.
\end{eqnarray*}

In our case, we use the following formula to calculate the Christoffel symbols:
\begin{eqnarray}
{\rm \Gamma}^A_{BC}&=&\frac12G^{AD}(\hat\partial_BG_{CD}+\hat\partial_CG_{BD}-\hat\partial_DG_{BC})
\nonumber\\
&&-\frac12G^{AD}(\mathbb C^E_{BD}G_{CE}+\mathbb C^E_{CD}G_{BE})+\frac12\mathbb C^A_{BC}.
\label{Christ_P}
\end{eqnarray}
In this formula, capital Latin indices mean  the condensed notations according to which, for example, $A=(i,a,\alpha)$.

As a result of the calculation performed by the formula (\ref{Christ_P}), we obtain the following Christoffel symbols:
\begin{eqnarray*}
&&{\rm \Gamma}^i_{jk}={}^{\scriptstyle\rm H}\tilde{\rm \Gamma}^i_{jk},\;\;\;{\rm \Gamma}^i_{jb}={}^{\scriptstyle\rm H}\tilde{\rm \Gamma}^i_{jb},\;\;\;{\rm \Gamma}^i_{ab}={}^{\scriptstyle\rm H}\tilde{\rm \Gamma}^i_{ab},
\nonumber\\
&&{\rm \Gamma}^a_{jk}={}^{\scriptstyle\rm H}\tilde{\rm \Gamma}^a_{jk},\;\;\;{\rm \Gamma}^a_{jb}={}^{\scriptstyle\rm H}\tilde{\rm \Gamma}^a_{jb},\;\;\;{\rm \Gamma}^a_{bc}={}^{\scriptstyle\rm H}\tilde{\rm \Gamma}^a_{bc},
\nonumber\\
&&{\rm \Gamma}^i_{j\alpha}=\frac12\tilde d_{\alpha\beta}(\tilde h^{im}{\tilde{\mathscr F}}^{\beta}_{jm}+\tilde h^{ia}{\tilde{\mathscr F}}^{\beta}_{ja}),\;\;\;\;{\rm \Gamma}^i_{\alpha j}={\rm \Gamma}^i_{j \alpha },
\nonumber\\
&&{\rm \Gamma}^i_{\alpha\beta}=-\frac12(\tilde h^{im}\tilde{\mathscr D}_m \tilde d_{\alpha\beta}+\tilde h^{ia}\tilde{\mathscr D}_a\tilde d_{\alpha\beta}),
\nonumber\\
&&\rm{where}\;\; \tilde{\mathscr D}_m \tilde d_{\alpha\beta}=\partial_m \tilde d_{\alpha\beta}-\tilde{\mathscr A}^{\sigma}_m(c^{\varepsilon}_{\sigma\alpha}\tilde d_{\varepsilon\beta}+c^{\varepsilon}_{\sigma\beta}\tilde d_{\varepsilon\alpha}).
\end{eqnarray*}
 
\begin{eqnarray*}
&&{\rm \Gamma}^a_{j\alpha}=\frac12\tilde d_{\alpha\beta}(\tilde h^{am}{\tilde{\mathscr F}}^{\beta}_{jm}+\tilde h^{ab}{\tilde{\mathscr F}}^{\beta}_{jb}),\;\;\;\;{\rm \Gamma}^a_{\alpha j}={\rm \Gamma}^a_{j\alpha},
\nonumber\\
&&{\rm \Gamma}^a_{b\mu}=\frac12\tilde d_{\mu\gamma}(\tilde h^{ac}{\tilde{\mathscr F}}^{\gamma}_{bc}+\tilde h^{an}{\tilde{\mathscr F}}^{\gamma}_{bn}),\;\;\;\;{\rm \Gamma}^a_{b\mu}={\rm \Gamma}^a_{\mu b},
\nonumber\\
&&{\rm \Gamma}^a_{\beta\alpha}=-\frac12\tilde h^{am}\hat H_m(\tilde d_{\beta\alpha})-\frac12 \tilde h^{aa'}\hat H_{a'}(\tilde d_{\beta\alpha}).
\end{eqnarray*}

\begin{eqnarray*}
&&{\rm \Gamma}^{\alpha}_{jk}=-\frac12{\tilde{\mathscr F}}^{\alpha}_{jk},\;\;\;{\rm \Gamma}^{\alpha}_{jb}=-\frac12{\tilde{\mathscr F}}^{\alpha}_{jb},\;\;\;{\rm \Gamma}^{\alpha}_{ab}=-\frac12{\tilde{\mathscr F}}^{\alpha}_{ab},
\nonumber\\
&&{\rm \Gamma}^{\alpha}_{\beta\gamma}=\frac12\tilde d^{\alpha\mu}(c^{\varepsilon}_{\beta\gamma}\tilde d_{\varepsilon\mu}-c^{\varepsilon}_{\mu\gamma}\tilde d_{\varepsilon\beta}-c^{\varepsilon}_{\mu\beta}\tilde d_{\varepsilon\gamma}),\;\;\;\;
\nonumber\\ 
&& {\rm \Gamma}^{\alpha}_{\alpha\gamma}=0 \,\,(\rm{for}\, \rm{the}\, \rm{semisiple}\, \rm{ Lie}\, \rm{group}),
\nonumber\\
&&{\rm \Gamma}^{\alpha}_{\beta k}=\frac12{\tilde d}^{\alpha\gamma}\hat H_k ({\tilde d}_{\beta\gamma}),\;\;\;{\rm \Gamma}^{\alpha}_{k \beta }={\rm \Gamma}^{\alpha}_{\beta k},
\nonumber\\
&&{\rm \Gamma}^{\alpha}_{\beta a}=\frac12{\tilde d}^{\alpha\gamma}\hat H_a ({\tilde d}_{\beta\gamma}),
\nonumber\\
&&{\rm \Gamma}^{\gamma}_{\gamma i}=\frac12 d^{\gamma \mu}{\partial}_i d_{\gamma \mu},\;\;\;{\rm \Gamma}^{\gamma}_{\gamma a}=\frac12 d^{\gamma \mu}{\partial}_a d_{\gamma \mu}.
\end{eqnarray*}

\appendix
\section*{Appendix B}
\section*{The scalar curvature of the manifold  $\tilde{\mathcal P}$}
\setcounter{equation}{0}
\def\theequation{B.\arabic{equation}}
In the article, the Riemann tensor  of the manifold $\tilde{\mathcal P}$ is defined by the Riemannian curvature operator, $\Omega(X,Y)=[\nabla_X,\nabla_Y]-\nabla_{[X,Y]}$ as follows:
$$\tilde R(X,Y,Z,Z')=G_{\tilde{\scriptstyle{\mathcal P}}}(\Omega(X,Y)Z,Z').$$
 The Ricci tensor  
$\tilde R_{AC}=\tilde R_{AMC}^{\;\;\;\;\;\;\;\;\;\;M}$ has the following representation in terms of Christoffel symbols:
$$\tilde R_{AC}=\hat\partial_A{\rm \Gamma}^B_{BC}-\hat\partial_B{\rm \Gamma}^B_{AC}+{\rm \Gamma}^D_{BC}{\rm \Gamma}^B_{AD}-{\rm \Gamma}^L_{AC}{\rm \Gamma}^B_{BL}-\mathbb C^E_{AB}{\rm \Gamma}^B_{EC}.$$
For tensors on $\tilde{\mathcal P}$, capital Latin letters, used as indices, denote three types of indices: $A=(i,a,\alpha)$.

Then the scalar curvature of the orbit space $\tilde{\mathcal M}$ is represented as
$$R_{\tilde{\mathcal M}}=\tilde h^{ik} R_{ik}+\tilde h^{ia} R_{ia}+\tilde h^{ai} R_{ai}+\tilde h^{ab} R_{ab}.$$
Here $R_{ik}$ is 
$$R_{ik}=\hat\partial_i{\rm \Gamma}^{B'}_{B'k}-\hat\partial_{B'}{\rm \Gamma}^{B'}_{ik}+{\rm \Gamma}^{D'}_{B'k}{\rm \Gamma}^{B'}_{iD'}-{\rm \Gamma}^{L'}_{ik}{\rm \Gamma}^{B'}_{B'L'}-\mathbb C^{E'}_{iB'}{\rm \Gamma}^{B'}_{E'k},$$
where capital letters with a prime as a superscript represent indices for tensors on $ \tilde{\mathcal M} $, and therefore they denote only two types of indices: $ B'= (i, b) $.
Note also that in our basis $(H_i,H_a,L_{\alpha})$,  the terms with $\mathbb C^{E'}_{iB'}$  will not contribute to $R_{ik}$.

The scalar curvature of the orbit $\mathcal G$ is obtained from the Ricci curvature $R_{\alpha\beta}$, which is defined as 
$$R_{\alpha\beta}=L_{\alpha}{\rm \Gamma}^{\gamma}_{\gamma\beta}-L_{\gamma}{\rm \Gamma}^{\gamma}_{\alpha\beta}+{\rm \Gamma}^{\mu}_{\nu\beta}{\rm \Gamma}^{\nu}_{\alpha\mu}-{\rm \Gamma}^{\sigma}_{\alpha\beta}{\rm \Gamma}^{\gamma}_{\gamma\sigma}-c^{\kappa}_{\alpha\gamma}{\rm \Gamma}^{\gamma}_{\kappa\beta}.$$
For the semisimple group Lie ${\rm \Gamma}^{\gamma}_{\gamma\beta}=0$, therefore we only have
$$R_{\alpha\beta}=-L_{\gamma}{\rm \Gamma}^{\gamma}_{\alpha\beta}+{\rm \Gamma}^{\mu}_{\nu\beta}{\rm \Gamma}^{\nu}_{\alpha\mu}-c^{\kappa}_{\alpha\gamma}{\rm \Gamma}^{\gamma}_{\kappa\beta}.$$
The scalar curvature of the orbit  $R_{\mathcal G}=\tilde d^{\alpha\beta} R_{\alpha\beta}$:
$$R_{\mathrm {\cal G}}=\frac12{d}^{\mu\nu} c^{\sigma}_{\mu \alpha} c^{\alpha}_{\nu\sigma}+
\frac14 {d}_{\mu\sigma}{d}^{\alpha\beta}{d}^{\epsilon\nu}
c^{\mu}_{\epsilon \alpha}c^{\sigma}_{\nu \beta}$$ is the scalar curvature of the orbit

\subsection*{Scalar curvature of $\tilde{\mathcal P}$}
The scalar curvature of $\tilde{\mathcal P}$ is defined as 
$$\tilde R_{\tilde{\mathcal P}}=\tilde h^{ik}\tilde R_{ik}+\tilde h^{ia} \tilde R_{ia}+\tilde h^{ai} \tilde R_{ai}+\tilde h^{ab} \tilde R_{ab}+\tilde d^{\alpha\beta} \tilde R_{\alpha\beta}.$$

The calculation of this scalar curvature is based on the assumption of its structure. Namely, we assume that $\tilde R_{\tilde{\mathcal P}}$ includes the scalar curvature $R_{\tilde{\mathcal M}}$, the scalar curvature $R_{\mathcal G}$,  the terms with the  square of the curvature of the mechanical connection, denoted as $FF$-term,  and the Laplace-Beltrami operator on  $R_{\tilde{\mathcal M}}$, acting on $\det d$. Given our assumption, we will successively analyze each term of  $\tilde R_{\tilde{\mathcal P}}$. We begin with the $\tilde R_{ik}$. 

The terms in $\tilde R_{ik}$ that remain after subtracting the terms used in the Ricci tensor   $R_{ik}$ of the manifold  $\tilde{\mathcal M}$ are as follows:
\begin{eqnarray}
&&-L_{\alpha} {\rm \Gamma}^{\alpha}_{ik}+\hat H_i{\rm \Gamma}^{\alpha}_{\alpha k}\nonumber\\
&&-{\rm \Gamma}^{\alpha}_{\alpha j}{\rm \Gamma}^j_{ik}-{\rm \Gamma}^{\alpha}_{\alpha b}{\rm \Gamma}^b_{ik}-{\rm \Gamma}^{b}_{b \alpha }{\rm \Gamma}^{\alpha}_{ik}-{\rm \Gamma}^{\beta}_{\beta\alpha}{\rm \Gamma}^{\alpha}_{ik}-{\rm \Gamma}^{n}_{n\alpha}{\rm \Gamma}^{\alpha}_{ik}\nonumber\\
&&+{\rm \Gamma}^{\beta}_{ij}{\rm \Gamma}^{j}_{\beta k}+{\rm \Gamma}^{\beta}_{ia}{\rm \Gamma}^{a}_{\beta k}+{\rm \Gamma}^{b}_{i \alpha}{\rm \Gamma}^{\alpha}_{bk}+{\rm \Gamma}^{\beta}_{i\alpha}{\rm \Gamma}^{\alpha}_{\beta k}+{\rm \Gamma}^{n}_{i\alpha}{\rm \Gamma}^{\alpha}_{nk}\nonumber\\
&&-\mathbb C^{\alpha}_{ij}{\rm \Gamma}^{j}_{\alpha k}-\mathbb C^{\alpha}_{ib}{\rm \Gamma}^{b}_{\alpha k}-\mathbb C^{\alpha}_{i\beta}{\rm \Gamma}^{\beta}_{\alpha k}
\label{R_ik_remain_terms}.
\end{eqnarray}
Thus, $\tilde R_{ik}=R_{ik}+$``$(\ref{R_ik_remain_terms})$''-terms.

Terms of the first line in (\ref{R_ik_remain_terms}):

$-L_{\alpha} {\rm \Gamma}^{\alpha}_{ik}=-L_{\alpha}(-\frac12\tilde{\mathscr F}^{\alpha}_{ik})=0$. This follows from 
$L_{\alpha}\bar{\rho}^{\alpha}_{\mu}=c^{\alpha}_{\sigma\alpha}\bar{\rho}^{\sigma}_{\mu}$, in which $c^{\alpha}_{\sigma\alpha}=0$, because in our case $\mathcal G$ is a     semisimple Lie group.

$\hat H_i{\rm \Gamma}^{\alpha}_{\alpha k}=\hat H_i(\frac12 \tilde d^{\alpha\beta}\hat H_k\tilde d_{\alpha\beta})=\frac12\partial_i(d^{\alpha\beta}\partial_k d_{\alpha\beta})$. Therefore in $\tilde R_{\tilde{\mathcal P}}$ we will have
$$\tilde h^{ik}\frac12\partial_i(d^{\alpha\beta}\partial_k d_{\alpha\beta})=\frac12(\tilde {h}^{ik}d^{\alpha\beta}\partial_i\partial_kd_{\alpha\beta}+\tilde {h}^{ik}(\partial_i d^{\alpha\beta})(\partial_k d_{\alpha\beta})).$$

The terms of the second line in (\ref{R_ik_remain_terms}):
\begin{eqnarray*}
&&-{\rm \Gamma}^{\alpha}_{\alpha j}{\rm \Gamma}^{j}_{ik}=-\frac12d^{\alpha\gamma}(\partial_jd_{\alpha\gamma}){\rm \Gamma}^{j}_{ik}
\nonumber\\
&&-{\rm \Gamma}^{\alpha}_{\alpha b}{\rm \Gamma}^{b}_{ik}=-\frac12d^{\alpha\gamma}(\partial_bd_{\alpha\gamma}){\rm \Gamma}^{b}_{ik}
\nonumber\\
&&-{\rm \Gamma}^{b}_{b\alpha }{\rm \Gamma}^{\alpha}_{ik}=\frac12d_{\alpha\gamma}\tilde h ^{bn}{\mathscr F}^{\gamma}_{bn}\cdot(-\frac12){\mathscr F}^{\alpha}_{ik}
\nonumber\\
&&-{\rm \Gamma}^{\beta}_{\beta\alpha }{\rm \Gamma}^{\alpha}_{ik}=0
\nonumber\\
&&-{\rm \Gamma}^{n}_{n\alpha }{\rm \Gamma}^{\alpha}_{ik}=-\frac12d_{\alpha\beta}(\tilde h^{nm}{\mathscr F}^{\beta}_{nm}+\tilde h ^{na}
{\mathscr F}^{\beta}_{na})\cdot(-\frac12{\mathscr F}^{\alpha}_{ik}).
\end{eqnarray*}

The terms of the third and fourth lines in (\ref{R_ik_remain_terms}):
\begin{eqnarray*}
&&{\rm \Gamma}^{\beta}_{ij}{\rm \Gamma}^{j}_{\beta k}=-\frac14d_{\beta\mu}(\tilde h^{jm}{\mathscr F}^{\beta}_{ij}{\mathscr F}^{\mu}_{km}+\tilde h^{ja}{\mathscr F}^{\beta}_{ij}{\mathscr F}^{\mu}_{ka})
\nonumber\\
&&{\rm \Gamma}^{\beta}_{ia}{\rm \Gamma}^{a}_{\beta k}=-\frac14d_{\beta\mu}(\tilde h^{am}{\mathscr F}^{\beta}_{ia}{\mathscr F}^{\mu}_{km}+\tilde h^{ab}{\mathscr F}^{\beta}_{ia}{\mathscr F}^{\mu}_{kb})
\nonumber\\
&&{\rm \Gamma}^{b}_{i \alpha}{\rm \Gamma}^{\alpha}_{bk}=-\frac14d_{\alpha\beta}(\tilde h^{bm}{\mathscr F}^{\beta}_{im}+\tilde h^{bc}{\mathscr F}^{\beta}_{ic}){\mathscr F}^{\alpha}_{bk}
\nonumber\\
&&{\rm \Gamma}^{\beta}_{i\alpha}{\rm \Gamma}^{\alpha}_{\beta k}=\frac14(\tilde d^{\beta\mu}\hat H_i\tilde d_{\alpha\mu})(\tilde d^{\alpha\nu}\hat H_k\tilde d_{\nu\beta})
\nonumber\\
&&{\rm \Gamma}^{n}_{i \alpha}{\rm \Gamma}^{\alpha}_{nk}=\frac12d_{\alpha\beta}(\tilde h^{nm}{\mathscr F}^{\beta}_{im}+\tilde h^{na}{\mathscr F}^{\beta}_{ia})\cdot (-\frac12){\mathscr F}^{\alpha}_{nk}
\nonumber\\
&&-\mathbb C^{\alpha}_{ij}{\rm \Gamma}^{j}_{\alpha k}=\frac12d_{\alpha\beta}(\tilde h^{jm}{\mathscr F}^{\beta}_{km}+\tilde h^{ja}{\mathscr F}^{\beta}_{ka}){\mathscr F}^{\alpha}_{ij}
\nonumber\\
&&\mathbb C^{\alpha}_{ib}{\rm \Gamma}^{b}_{\alpha k}=\frac12d_{\alpha\beta}(\tilde h^{bm}{\mathscr F}^{\beta}_{km}+\tilde h^{bc}{\mathscr F}^{\beta}_{kc}){\mathscr F}^{\alpha}_{ib}
\nonumber\\
&&\mathbb C^{\alpha}_{i\beta}{\rm \Gamma}^{\beta}_{\alpha k}=0.
\end{eqnarray*}

The remaining terms in $\tilde R_{ia}$:
\begin{eqnarray}
&&-L_{\alpha} {\rm \Gamma}^{\alpha}_{ia}+\hat H_i{\rm \Gamma}^{\gamma}_{\gamma a}\nonumber\\
&&-{\rm \Gamma}^{\gamma}_{\gamma j}{\rm \Gamma}^j_{ia}-{\rm \Gamma}^{\gamma}_{\gamma b}{\rm \Gamma}^b_{ia}-{\rm \Gamma}^{\mu}_{\mu\gamma }{\rm \Gamma}^{\gamma}_{ia}-{\rm \Gamma}^{n}_{n\gamma }{\rm \Gamma}^{\gamma}_{ia}-{\rm \Gamma}^{b}_{b\gamma }{\rm \Gamma}^{\gamma}_{ia}\nonumber\\
&&+{\rm \Gamma}^{\beta}_{in}{\rm \Gamma}^{n}_{\beta a}+{\rm \Gamma}^{\mu}_{ib}{\rm \Gamma}^{b}_{\mu a}+{\rm \Gamma}^{n}_{i \beta}{\rm \Gamma}^{\beta}_{na}+{\rm \Gamma}^{c}_{i\alpha}{\rm \Gamma}^{\alpha}_{ca}+{\rm \Gamma}^{\mu}_{i\beta }{\rm \Gamma}^{\beta}_{\mu a}\nonumber\\
&&-\mathbb C^{\alpha}_{in}{\rm \Gamma}^{n}_{\alpha a}-\mathbb C^{\alpha}_{ib}{\rm \Gamma}^{b}_{\alpha a}-\mathbb C^{\alpha}_{i\beta}{\rm \Gamma}^{\beta}_{\alpha a}.
\label{R_ia_remain_terms}
\end{eqnarray}
Thus, $\tilde R_{ia}=R_{ia}+$``$(\ref{R_ia_remain_terms})$''-terms.

Terms of the first line in (\ref{R_ia_remain_terms}):

$-L_{\alpha}{\rm \Gamma}^{\alpha}_{ia}=0,$

$$\hat H_i{\rm \Gamma}^{\gamma}_{\gamma a}=\hat H_i(\frac12\tilde d^{\gamma\mu}(\hat H_a \tilde d_{\gamma\mu}))=\frac12(d^{\gamma\mu}\partial_i\partial_a d_{\gamma\mu}+(\partial_i d^{\gamma\mu})(\partial_a d_{\gamma\mu})).$$

Terms of the second line in (\ref{R_ia_remain_terms}):
\begin{eqnarray*}
&&{\rm \Gamma}^{\gamma}_{\gamma j}{\rm \Gamma}^{j}_{i a}=-\frac12d^{\gamma\mu}(\hat H_jd_{\gamma\mu}){\rm \Gamma}^{j}_{i a}
\nonumber\\
&&{\rm \Gamma}^{\gamma}_{\gamma b}{\rm \Gamma}^{b}_{i a}=-\frac12d^{\gamma\mu}(\hat H_bd_{\gamma\mu}){\rm \Gamma}^{b}_{i a}
\nonumber\\
&&{\rm \Gamma}^{n}_{n \gamma }{\rm \Gamma}^{\gamma}_{i a}=-\frac12\tilde d_{\gamma\beta}(\tilde h^{nm}\tilde{\mathscr F}^{\beta}_{nm}+\tilde h^{nb}\tilde{\mathscr F}^{\beta}_{nb})\Bigl(-\frac12\tilde{\mathscr F}^{\gamma}_{ia}\Bigr)
\nonumber\\
&&{\rm \Gamma}^{b}_{b \gamma }{\rm \Gamma}^{\gamma}_{i a}=-\frac12\tilde d_{\gamma\mu}(\tilde h^{bc}\tilde{\mathscr F}^{\mu}_{bc}+\tilde h^{bm}\tilde{\mathscr F}^{\mu}_{bm})\Bigl(-\frac12\tilde{\mathscr F}^{\gamma}_{ia}\Bigr)
\nonumber\\
&&{\rm \Gamma}^{\mu}_{\mu \gamma }{\rm \Gamma}^{\gamma}_{i a}=0.
\end{eqnarray*}

The terms of the third and fourth lines  in (\ref{R_ia_remain_terms}):
\begin{eqnarray*}
&&{\rm \Gamma}^{\beta}_{in}{\rm \Gamma}^{n}_{\beta a}=-(\frac12 \tilde {\mathscr F}^{\alpha}_{in})\cdot \frac12\tilde d_{\alpha\gamma}(\tilde h^{nm}\tilde{\mathscr F}^{\gamma}_{am}+\tilde h^{nb}\tilde{\mathscr F}^{\gamma
}_{ab})
\nonumber\\
&&{\rm \Gamma}^{\mu}_{ib}{\rm \Gamma}^{b}_{\mu a}=(-\frac12{\mathscr F}^{\mu}_{ib})\cdot\frac12d_{\mu\nu}(\tilde h^{bm}{\mathscr F}^{\nu}_{am}+\tilde h^{bc}{\mathscr F}^{\nu}_{ac})
\nonumber\\
&&{\rm \Gamma}^{n}_{i \beta}{\rm \Gamma}^{\beta}_{na}=\frac12\tilde d_{\beta\gamma}(\tilde h^{nm}\tilde {\mathscr F}^{\gamma}_{im}+\tilde h^{nb}\tilde {\mathscr F}^{\gamma}_{ib})(-\frac12\tilde {\mathscr F}^{\beta}_{na})
\nonumber\\
&&{\rm \Gamma}^{c}_{i\alpha}{\rm \Gamma}^{\alpha}_{ca}=\frac12\tilde d_{\alpha\beta}(\tilde h^{cm}\tilde {\mathscr F}^{\beta}_{im}+\tilde h^{cb}\tilde {\mathscr F}^{\beta}_{ib})(-\frac12\tilde {\mathscr F}^{\alpha}_{ca})
\nonumber\\
&&{\rm \Gamma}^{\mu}_{i\beta }{\rm \Gamma}^{\beta}_{\mu a}=\frac14(\tilde d^{\mu\gamma}\hat H_i\tilde d_{\gamma\alpha})(\tilde d^{\alpha\nu}\hat H_a\tilde d_{\mu\nu})
\nonumber\\
&&-\mathbb C^{\alpha}_{in}{\rm \Gamma}^{n}_{\alpha a}=-(-{\mathscr F}^{\alpha}_{in})\cdot\frac12d_{\alpha\beta}(\tilde h^{nm}{\mathscr F}^{\beta}_{am}+\tilde h^{nb}{\mathscr F}^{\beta}_{ab})
\nonumber\\
&&-\mathbb C^{\alpha}_{ib}{\rm \Gamma}^{b}_{\alpha a}=-(-{\mathscr F}^{\alpha}_{ib})\cdot\frac12d_{\alpha\gamma}(\tilde h^{bc}{\mathscr F}^{\gamma}_{ac}+\tilde h^{bm}{\mathscr F}^{\gamma}_{am})
\nonumber\\
&&\mathbb C^{\alpha}_{i\beta}{\rm \Gamma}^{\beta}_{\alpha a}=0.
\end{eqnarray*}

The remaining terms in $\tilde R_{ai}$:
\begin{eqnarray}
&&-L_{\beta} {\rm \Gamma}^{\beta}_{ai}+\hat H_a{\rm \Gamma}^{\beta}_{\beta i}\nonumber\\
&&-{\rm \Gamma}^{\gamma}_{\gamma n}{\rm \Gamma}^n_{ai}-{\rm \Gamma}^{\gamma}_{\gamma b}{\rm \Gamma}^b_{ai}-{\rm \Gamma}^{\mu}_{\mu\gamma }{\rm \Gamma}^{\gamma}_{ai}-{\rm \Gamma}^{n}_{n\gamma }{\rm \Gamma}^{\gamma}_{ai}-{\rm \Gamma}^{b}_{b\alpha }{\rm \Gamma}^{\alpha}_{ai}\nonumber\\
&&+{\rm \Gamma}^{\beta}_{an}{\rm \Gamma}^{n}_{\beta i}+{\rm \Gamma}^{\mu}_{ab}{\rm \Gamma}^{b}_{\mu i}+{\rm \Gamma}^{n}_{a \beta}{\rm \Gamma}^{\beta}_{ni}+{\rm \Gamma}^{b}_{a\beta}{\rm \Gamma}^{\beta}_{bi}+{\rm \Gamma}^{\mu}_{i\beta }{\rm \Gamma}^{\beta}_{\mu a}\nonumber\\
&&-\mathbb C^{\alpha}_{an}{\rm \Gamma}^{n}_{\alpha i}-\mathbb C^{\alpha}_{ab}{\rm \Gamma}^{b}_{\alpha i}-\mathbb C^{\alpha}_{a\gamma}{\rm \Gamma}^{\gamma}_{\alpha i}.
\label{R_ai_remain_terms}
\end{eqnarray}
 $\tilde R_{ai}=R_{ai}+$``$(\ref{R_ai_remain_terms})$''-terms.

Terms of the first line in (\ref{R_ai_remain_terms}):
$-L_{\beta}{\rm \Gamma}^{\beta}_{ai}=0,$

$$\hat H_a{\rm \Gamma}^{\beta}_{\beta i}=\hat H_a(\frac12\tilde d^{\alpha\gamma}(\hat H_i \tilde d_{\alpha\gamma}))=\frac12(d^{\alpha\gamma}\partial_a\partial_i d_{\alpha\gamma}+(\partial_a d^{\alpha\gamma})(\partial_i d_{\alpha\gamma})).$$

Terms of the second line in (\ref{R_ai_remain_terms}):
\begin{eqnarray*}
&&-{\rm \Gamma}^{\gamma}_{\gamma n}{\rm \Gamma}^n_{ai}=-\frac12(d^{\alpha\gamma}\partial_n d_{\alpha\gamma}){\rm \Gamma}^n_{ai}
\nonumber\\
&&-{\rm \Gamma}^{\gamma}_{\gamma b}{\rm \Gamma}^b_{ai}=-\frac12(d^{\alpha\gamma}\partial_b d_{\alpha\gamma}){\rm \Gamma}^b_{ai}
\nonumber\\
&&-{\rm \Gamma}^{\mu}_{\mu\gamma }{\rm \Gamma}^{\gamma}_{ai}=0
\nonumber\\
&&-{\rm \Gamma}^{n}_{n\gamma }{\rm \Gamma}^{\gamma}_{ai}=-\frac12\tilde d_{\gamma\beta}(\tilde h^{nm}\tilde{\mathscr F}^{\beta}_{nm}+\tilde h^{na}\tilde{\mathscr F}^{\beta}_{na}){\rm \Gamma}^{\gamma}_{ai}
\nonumber\\
&&-{\rm \Gamma}^{b}_{b\alpha }{\rm \Gamma}^{\alpha}_{ai}=-\frac12\tilde d_{\alpha\beta}(\tilde h^{bc}\tilde{\mathscr F}^{\beta}_{bc}+\tilde h^{bn}\tilde{\mathscr F}^{\beta}_{bn}){\rm \Gamma}^{\alpha}_{ai}.
\end{eqnarray*}

Terms of the fourth and fifth  lines in (\ref{R_ai_remain_terms}):
\begin{eqnarray*}
&&{\rm \Gamma}^{\alpha}_{an}{\rm \Gamma}^{n}_{\alpha i}=(-\frac12 \tilde {\mathscr F}^{\alpha}_{an})\cdot \frac12\tilde d_{\alpha\gamma}(\tilde h^{nm}\tilde {\mathscr F}^{\gamma}_{im}+\tilde h^{nb}\tilde {\mathscr F}^{\gamma}_{ib})\nonumber\\
&&{\rm \Gamma}^{\mu}_{ab}{\rm \Gamma}^{b}_{\mu i}=(-\frac12{\mathscr F}^{\mu}_{ab})\cdot\frac12d_{\mu\nu}(\tilde h^{bm}{\mathscr F}^{\nu}_{im}+\tilde h^{bc}{\mathscr F}^{\nu}_{ic})\nonumber\\
&&{\rm \Gamma}^{n}_{a \beta}{\rm \Gamma}^{\beta}_{ni}=\frac12\tilde d_{\beta\gamma}(\tilde h^{nm}\tilde {\mathscr F}^{\gamma}_{am}+\tilde h^{nb}\tilde {\mathscr F}^{\gamma}_{ab})(-\frac12\tilde {\mathscr F}^{\beta}_{ni})\nonumber\\
&&{\rm \Gamma}^{b}_{a\beta}{\rm \Gamma}^{\beta}_{bi}=\frac12\tilde d_{\beta\gamma}(\tilde h^{bc}\tilde {\mathscr F}^{\gamma}_{ac}+\tilde h^{bm}\tilde {\mathscr F}^{\gamma}_{am})(-\frac12\tilde {\mathscr F}^{\beta}_{bi})\nonumber\\
&&{\rm \Gamma}^{\mu}_{i\beta }{\rm \Gamma}^{\beta}_{\mu a}=\frac14(\tilde d^{\mu\gamma}\hat H_i\tilde d_{\gamma\beta})(\tilde d^{\beta\nu}\hat H_a\tilde d_{\mu\nu})\nonumber\\
&&-\mathbb C^{\alpha}_{an}{\rm \Gamma}^{n}_{\alpha i}=-(-{\mathscr F}^{\alpha}_{an})\cdot\frac12d_{\alpha\beta}(\tilde h^{nm}{\mathscr F}^{\beta}_{im}+\tilde h^{nb}{\mathscr F}^{\beta}_{ib})\nonumber\\
&&-\mathbb C^{\alpha}_{ab}{\rm \Gamma}^{b}_{\alpha i}=-(-{\mathscr F}^{\alpha}_{ab})\cdot\frac12d_{\alpha\gamma}(\tilde h^{bm}{\mathscr F}^{\gamma}_{im}+\tilde h^{ba}{\mathscr F}^{\gamma}_{ia})\nonumber\\
&&-\mathbb C^{\alpha}_{a\gamma}{\rm \Gamma}^{\gamma}_{\alpha i}=0.\nonumber\\
\end{eqnarray*}

The remaining terms in $\tilde R_{ab}$:
\begin{eqnarray}
&&-L_{\alpha} {\rm \Gamma}^{\alpha}_{ab}+\hat H_a{\rm \Gamma}^{\mu}_{\mu b}\nonumber\\
&&-{\rm \Gamma}^{\mu}_{\mu n}{\rm \Gamma}^n_{ab}-{\rm \Gamma}^{\mu}_{\mu c}{\rm \Gamma}^c_{ab}-{\rm \Gamma}^{n}_{n\alpha }{\rm \Gamma}^{\alpha}_{ab}-{\rm \Gamma}^{c}_{c\alpha }{\rm \Gamma}^{\alpha}_{ab}-{\rm \Gamma}^{\mu}_{\mu\alpha }{\rm \Gamma}^{\alpha}_{ab}
\nonumber\\
&&+{\rm \Gamma}^{\alpha}_{an}{\rm \Gamma}^{n}_{\alpha b}+{\rm \Gamma}^{\alpha}_{ac}{\rm \Gamma}^{c}_{\alpha b}+{\rm \Gamma}^{m}_{a \beta}{\rm \Gamma}^{\beta}_{mb}+{\rm \Gamma}^{c}_{a\beta}{\rm \Gamma}^{\beta}_{cb}+{\rm \Gamma}^{\beta}_{a\mu }{\rm \Gamma}^{\mu}_{\beta b}\nonumber\\
&&-\mathbb C^{\alpha}_{ai}{\rm \Gamma}^{i}_{\alpha b}-\mathbb C^{\alpha}_{ac}{\rm \Gamma}^{c}_{\alpha b}-\mathbb C^{\alpha}_{a \mu}{\rm \Gamma}^{\mu}_{\alpha b}.
\label{R_ab_remain_terms}
\end{eqnarray}
Thus, $\tilde R_{ab}=R_{ab}+$``$(\ref{R_ab_remain_terms})$''-terms.

Terms of the first line in (\ref{R_ab_remain_terms}):

$-L_{\alpha}{\rm \Gamma}^{\alpha}_{ab}=0,$
$$\hat H_a{\rm \Gamma}^{\mu}_{\mu b}=\hat H_a(\frac12\tilde d^{\mu\gamma}(\hat H_b \tilde d_{\mu\gamma}))=\frac12(d^{\mu\gamma}\partial_a\partial_b d_{\mu\gamma}+(\partial_a d^{\mu\gamma})(\partial_b d_{\mu\gamma})).$$

Terms of the second line in (\ref{R_ab_remain_terms}):
$$-{\rm \Gamma}^{\mu}_{\mu n}{\rm \Gamma}^{n}_{a b}=-\frac12d^{\mu\gamma}(\partial_n d_{\mu\gamma}){\rm \Gamma}^{n}_{a b}$$

$$-{\rm \Gamma}^{\mu}_{\mu c}{\rm \Gamma}^{c}_{a b}=-\frac12d^{\mu\gamma}(\partial_c d_{\mu\gamma}){\rm \Gamma}^{c}_{a b}$$

$$-{\rm \Gamma}^{n}_{n \alpha}{\rm \Gamma}^{\alpha}_{a b}=-\frac12\tilde d_{\alpha\beta}(\tilde h^{nm}\tilde {\mathscr F}^{\beta}_{nm}+\tilde h^{na}\tilde {\mathscr F}^{\beta}_{na})(-\tilde {\mathscr F}^{\alpha}_{ab}),$$
where the first term on the right vanishes.

$$-{\rm \Gamma}^{c}_{c \alpha}{\rm \Gamma}^{\alpha}_{a b}=-\frac12\tilde d_{\alpha\beta}(\tilde h^{cm}\tilde {\mathscr F}^{\beta}_{cm}+\tilde h^{cd}\tilde {\mathscr F}^{\beta}_{cd})(-\tilde {\mathscr F}^{\alpha}_{ab}),$$
where the second term on the right vanishes.

$$-{\rm \Gamma}^{\mu}_{\mu \alpha}{\rm \Gamma}^{\alpha}_{a b}=0,$$
since ${\rm \Gamma}^{\mu}_{\mu \alpha}=0$ for the semisimple Lie group.

Terms of the third and fourth  lines in (\ref{R_ab_remain_terms}):
$${\rm \Gamma}^{\alpha}_{an}{\rm \Gamma}^{n}_{\alpha b}=(-\frac12 \tilde {\mathscr F}^{\alpha}_{an})\cdot \frac12\tilde d_{\alpha\gamma}(\tilde h^{nm}\tilde {\mathscr F}^{\gamma}_{bm}+\tilde h^{nc}\tilde {\mathscr F}^{\gamma}_{bc})$$

$${\rm \Gamma}^{\alpha}_{ac}{\rm \Gamma}^{c}_{\alpha b}=(-\frac12{\mathscr F}^{\alpha}_{ac})\cdot\frac12d_{\alpha\gamma}(\tilde h^{cd}{\mathscr F}^{\gamma}_{bd}+\tilde h^{cm}{\mathscr F}^{\gamma}_{bm})$$

$${\rm \Gamma}^{m}_{a \beta}{\rm \Gamma}^{\beta}_{mb}=\frac12\tilde d_{\beta\gamma}(\tilde h^{mk}\tilde {\mathscr F}^{\gamma}_{ak}+\tilde h^{mc}\tilde {\mathscr F}^{\gamma}_{ac})(-\frac12\tilde {\mathscr F}^{\beta}_{mb})$$

$${\rm \Gamma}^{c}_{a\alpha}{\rm \Gamma}^{\alpha}_{c b}=\frac12\tilde d_{\alpha\gamma}(\tilde h^{cd}\tilde {\mathscr F}^{\gamma}_{ad}+\tilde h^{cm}\tilde {\mathscr F}^{\gamma}_{am})(-\frac12\tilde {\mathscr F}^{\alpha}_{cb})$$

$${\rm \Gamma}^{\beta}_{a\mu }{\rm \Gamma}^{\mu}_{\beta b}=\frac14(\tilde d^{\beta\gamma}\hat H_a\tilde d_{\gamma\mu})(\tilde d^{\mu\nu}\hat H_b\tilde d_{\nu\beta})$$

$$-\mathbb C^{\alpha}_{ai}{\rm \Gamma}^{i}_{\alpha b}=-(-{\mathscr F}^{\alpha}_{ai})\cdot\frac12d_{\alpha\beta}(\tilde h^{im}{\mathscr F}^{\beta}_{bm}+\tilde h^{ic}{\mathscr F}^{\beta}_{bc})$$

$$-\mathbb C^{\alpha}_{ac}{\rm \Gamma}^{c}_{\alpha c}=-(-{\mathscr F}^{\alpha}_{ac})\cdot\frac12d_{\alpha\gamma}(\tilde h^{cm}{\mathscr F}^{\gamma}_{bm}+\tilde h^{cd}{\mathscr F}^{\gamma}_{bd})$$

$$-\mathbb C^{\alpha}_{a\gamma}{\rm \Gamma}^{\gamma}_{\alpha b}=0.$$

The remaining terms in $\tilde R_{\alpha\beta}$:

\begin{eqnarray}
&&-\hat H_{i} {\rm \Gamma}^i_{\alpha\beta}-\hat H_a{\rm \Gamma}^a_{\alpha\beta}+L_{\alpha}{\rm \Gamma}^i_{i\beta}+L_{\alpha}{\rm \Gamma}^a_{a\beta}\nonumber\\
&&-({\rm \Gamma}^{k}_{k i}+{\rm \Gamma}^a_{ai}+{\rm \Gamma}^{\gamma}_{\gamma i}){\rm \Gamma}^i_{\alpha\beta}-({\rm \Gamma}^{k}_{ka}+{\rm \Gamma}^{b}_{ba}+{\rm \Gamma}^{\gamma}_{\gamma a }){\rm \Gamma}^a_{\alpha \beta}-{\rm \Gamma}^i_{i \gamma}{\rm \Gamma}^{\gamma}_{\alpha\beta}-{\rm \Gamma}^a_{a \gamma}{\rm \Gamma}^{\gamma}_{\alpha\beta}\nonumber\\
&&+{\rm \Gamma}^j_{\alpha i}{\rm \Gamma}^{i}_{j\beta}+{\rm \Gamma}^a_{\alpha i}{\rm \Gamma}^{i}_{a\beta}+{\rm \Gamma}^{\mu}_{\alpha i }{\rm \Gamma}^i_{\mu\beta}+{\rm \Gamma}^{i}_{\alpha b}{\rm \Gamma}^b_{i\beta}+{\rm \Gamma}^a_{\alpha b}{\rm \Gamma}^b_{a\beta}+{\rm \Gamma}^{\mu}_{\alpha b }{\rm \Gamma}^b_{\mu\beta}+{\rm \Gamma}^{i}_{\alpha \gamma}{\rm \Gamma}^{\gamma}_{i\beta}+{\rm \Gamma}^{a}_{\alpha \gamma }{\rm \Gamma}^{\gamma}_{a\beta}\nonumber\\
&&-\mathbb C^{\gamma}_{\alpha i}{\rm \Gamma}^{i}_{\gamma \beta}-\mathbb C^{\gamma}_{\alpha b}{\rm \Gamma}^{b}_{\gamma \beta}
\label{R_alfabeta_remain_terms})
\end{eqnarray}
 $\tilde R_{\alpha\beta}=R_{\alpha\beta}+$``$(\ref{R_alfabeta_remain_terms})$''-terms.

Terms of the first line in (\ref{R_alfabeta_remain_terms}):
\begin{eqnarray*}
&&-\hat H_{i} {\rm \Gamma}^i_{\alpha\beta}=-\hat H_i\,\bigl[-\frac12(\tilde h ^{im}\tilde{\mathscr D}_m \tilde d_{\alpha\beta}+\tilde h ^{ia}\tilde{\mathscr D}_a \tilde d_{\alpha\beta})\bigr],
\nonumber\\
&&\hat H_a{\rm \Gamma}^a_{\alpha\beta}=\hat H_a\bigl[\frac12(\tilde h^{am}\hat H_m\tilde d_{\alpha\beta}+\tilde h^{ab}\hat H_b\tilde d_{\alpha\beta})\bigr],
\nonumber\\
&&L_{\alpha}{\rm \Gamma}^i_{i\beta}=L_{\alpha}\bigl[\frac12\tilde d_{\beta\gamma}(\tilde h^{im}\tilde{\mathscr F}^{\gamma}_{im}+\tilde h^{ia}\tilde{\mathscr F}^{\gamma}_{ia})\bigr]=\frac12L_{\alpha}\bigl[\tilde d_{\beta\gamma}\,\tilde h^{ia}\tilde{\mathscr F}^{\gamma}_{ia}\bigr],
\nonumber\\
&&L_{\alpha}{\rm \Gamma}^a_{a\beta}=L_{\alpha}\bigl[\frac12\tilde d_{\beta\gamma}(\tilde h^{ac}\tilde{\mathscr F}^{\gamma}_{ac}+\tilde h^{an}\tilde{\mathscr F}^{\gamma}_{an})\bigr]=\frac12L_{\alpha}\bigl[\tilde d_{\beta\gamma}\,\tilde h^{an}\tilde{\mathscr F}^{\gamma}_{an}\bigr].
\end{eqnarray*}
Terms of the second line in (\ref{R_alfabeta_remain_terms}):
\begin{eqnarray*}
&&-({\rm \Gamma}^{k}_{k i}+{\rm \Gamma}^a_{ai}+{\rm \Gamma}^{\gamma}_{\gamma i}){\rm \Gamma}^i_{\alpha\beta}=\frac12({\rm \Gamma}^{k}_{k i}+{\rm \Gamma}^a_{ai}+{\rm \Gamma}^{\gamma}_{\gamma i})(\tilde h ^{im}\tilde{\mathscr D}_m \tilde d_{\alpha\beta}+\tilde h ^{ia}\tilde{\mathscr D}_a \tilde d_{\alpha\beta})
\nonumber\\
&&-({\rm \Gamma}^{k}_{ka}+{\rm \Gamma}^{b}_{ba}+{\rm \Gamma}^{\gamma}_{\gamma a }){\rm \Gamma}^a_{\alpha \beta}=\frac12({\rm \Gamma}^{k}_{ka}+{\rm \Gamma}^{b}_{ba}+{\rm \Gamma}^{\gamma}_{\gamma a })(\tilde h^{am}\hat H_m\tilde d_{\alpha\beta}+\tilde h^{ab}\hat H_b\tilde d_{\alpha\beta})
\nonumber\\
&&-({\rm \Gamma}^i_{i \gamma}+{\rm \Gamma}^a_{a \gamma}){\rm \Gamma}^{\gamma}_{\alpha\beta}=-\frac12({\rm \Gamma}^i_{i \gamma}+{\rm \Gamma}^a_{a \gamma})(c^{\gamma}_{\alpha\beta}-\tilde d^{\gamma\sigma}c^{\varphi}_{\sigma\beta}\tilde d_{\varphi\alpha}-\tilde d^{\gamma\sigma}c^{\varphi}_{\sigma\alpha}\tilde d_{\varphi\beta})
\nonumber\\
&&{\rm \Gamma}^{\gamma}_{\gamma a }=\frac12(d^{\alpha\mu}\partial_ad_{\alpha\mu}),\;\;\;{\rm \Gamma}^{\gamma}_{\gamma i}=\frac12(d^{\alpha\mu}\partial_id_{\alpha\mu}),
\nonumber\\
&&{\rm \Gamma}^i_{i \gamma}=\frac12(d_{\beta\gamma}\,\tilde h^{ia}\tilde{\mathscr F}^{\beta}_{ia}),\;\;\;{\rm \Gamma}^a_{a \gamma}=\frac12(d_{\mu\gamma}\,\tilde h^{an}\tilde{\mathscr F}^{\mu}_{an}).
\end{eqnarray*}

Terms of the third line in (\ref{R_alfabeta_remain_terms}):
\begin{eqnarray*}
&&{\rm \Gamma}^j_{\alpha i}{\rm \Gamma}^{i}_{j\beta}=\frac12\tilde d_{\alpha\nu}(\tilde h^{jn}\tilde{\mathscr F}^{\nu}_{in}+\tilde h^{jb}\tilde{\mathscr F}^{\nu}_{ib})\cdot\frac12\tilde d_{\beta\mu}(\tilde h^{im}\tilde{\mathscr F}^{\mu}_{jm}+\tilde h^{ia}\tilde{\mathscr F}^{\mu}_{ja})
\nonumber\\
&&{\rm \Gamma}^a_{\alpha i}{\rm \Gamma}^{i}_{a\beta}=\frac12\tilde d_{\alpha\mu}(\tilde h^{am}\tilde{\mathscr F}^{\mu}_{im}+\tilde h^{ab}\tilde{\mathscr F}^{\mu}_{ib})\cdot\frac12\tilde d_{\beta\nu}(\tilde h^{ik}\tilde{\mathscr F}^{\nu}_{ak}+\tilde h^{ic}\tilde{\mathscr F}^{\nu}_{ac})
\nonumber\\
&&{\rm \Gamma}^a_{\alpha i}{\rm \Gamma}^{i}_{a\beta}=\frac12\tilde d_{\alpha\mu}(\tilde h^{am}\tilde{\mathscr F}^{\mu}_{im}+\tilde h^{ab}\tilde{\mathscr F}^{\mu}_{ib})\cdot\frac12\tilde d_{\beta\nu}(\tilde h^{ik}\tilde{\mathscr F}^{\nu}_{ak}+\tilde h^{ic}\tilde{\mathscr F}^{\nu}_{acb})
\nonumber\\
&&{\rm \Gamma}^{\mu}_{\alpha i }{\rm \Gamma}^i_{\mu\beta}=\frac12 \tilde d^{\mu\gamma}(\hat H_i\tilde d_{\gamma\alpha})\cdot(-\frac12)(\tilde h ^{im}\tilde{\mathscr D}_m \tilde d_{\mu\beta}+\tilde h ^{ia}\tilde{\mathscr D}_a \tilde d_{\mu\beta})
\nonumber\\
&&{\rm \Gamma}^{i}_{\alpha b}{\rm \Gamma}^b_{i\beta}=\frac12\tilde d_{\alpha\gamma}(\tilde h^{im}\tilde{\mathscr F}^{\gamma}_{bm}+\tilde h^{ia}\tilde{\mathscr F}^{\gamma}_{ba})\cdot\frac12\tilde d_{\beta\nu}(\tilde h^{bk}\tilde{\mathscr F}^{\nu}_{ik}+\tilde h^{bc}\tilde{\mathscr F}^{\nu}_{ic})
\nonumber\\
&&{\rm \Gamma}^a_{\alpha b}{\rm \Gamma}^b_{a\beta}=\frac12\tilde d_{\alpha\gamma}(\tilde h^{ac}\tilde{\mathscr F}^{\gamma}_{bc}+\tilde h^{am}\tilde{\mathscr F}^{\gamma}_{bm})\cdot\frac12\tilde d_{\beta\mu}(\tilde h^{bd}\tilde{\mathscr F}^{\mu}_{ad}+\tilde h^{bk}\tilde{\mathscr F}^{\mu}_{ak})
\nonumber\\
&&{\rm \Gamma}^{\mu}_{\alpha b }{\rm \Gamma}^b_{\mu\beta}=\frac12 \tilde d^{\mu\gamma}(\hat H_b\tilde d_{\alpha\gamma})\cdot(-\frac12)(\tilde h^{bk}\hat H_k\tilde d_{\mu\beta}+\tilde h^{bc}\hat H_c\tilde d_{\mu\beta})
\nonumber\\
&&{\rm \Gamma}^{i}_{\alpha \gamma}{\rm \Gamma}^{\gamma}_{i\beta}=-\frac12(\tilde h ^{im}\tilde{\mathscr D}_m \tilde d_{\alpha\gamma}+\tilde h ^{ia}\tilde{\mathscr D}_a \tilde d_{\alpha\gamma})\cdot\frac12\tilde d^{\gamma\sigma}(\hat H_i\tilde d_{\beta\sigma})
\nonumber\\
&&{\rm \Gamma}^{a}_{\alpha \gamma }{\rm \Gamma}^{\gamma}_{a\beta}=-\frac12(\tilde h^{am}\hat H_m\tilde d_{\alpha\gamma}+\tilde h^{ab}\hat H_b\tilde d_{\alpha\gamma})\cdot \frac12 d^{\gamma\mu}(\hat H_ad_{\mu\beta})
\nonumber\\
&&-\mathbb C^{\gamma}_{\alpha i}{\rm \Gamma}^{i}_{\gamma \beta}=0,\;\;\;\;-\mathbb C^{\gamma}_{\alpha b}{\rm \Gamma}^{b}_{\gamma \beta}=0.
\end{eqnarray*}

To obtain the expression for the scalar curvature $\tilde R$, we first arrange  the elements of (\ref{R_ik_remain_terms}), (\ref{R_ia_remain_terms}), (\ref{R_ai_remain_terms}), (\ref{R_ab_remain_terms}) and (\ref{R_alfabeta_remain_terms}) in four group. Also at this stage of the consideration, we do not take into account terms that depend on $\mathscr F$.

 In the first group we include 2, 3, 4, 11 terms  of (\ref{R_ik_remain_terms}) and those parts of 1, 5, 6, 7, 15, 19 terms of (\ref{R_alfabeta_remain_terms}) that contain covariant (or partial) derivatives only with respect to the variables $x^i$.

 In $\tilde R$, terms belonging to (\ref{R_ik_remain_terms}) that were taken to form the first group are represented as follows:

$$\hat H_i\Gamma^{\alpha}_{\alpha k}\tilde h^{ik}=\frac12\tilde h^{ik}\partial_i(d^{\alpha\beta}\partial_kd_{\alpha\beta}),$$

$$-\Gamma^{\alpha}_{\alpha j}\Gamma^{j}_{ik}\tilde h^{ik}=-\frac12\tilde h^{ik}\Gamma^{j}_{ik}(d^{\alpha\gamma}\partial_jd_{\alpha\gamma}),$$

$$-\Gamma^{\alpha}_{\alpha b}\Gamma^{b}_{ik}\tilde h^{ik}=-\frac12\tilde h^{ik}\Gamma^{b}_{ik}(d^{\alpha\gamma}\partial_bd_{\alpha\gamma}),$$

$$\Gamma^{\beta}_{i\alpha }\Gamma^{\alpha}_{\beta k}\tilde h^{ik}=\frac14\tilde h^{ik}(d^{\beta\mu}\hat H_id_{\alpha\mu})(d^{\alpha\nu}\hat H_kd_{\nu\beta})=\frac14\tilde h^{ik}(d^{\beta\mu}{\mathscr D}_id_{\alpha\mu})(d^{\alpha\nu}{\mathscr D}_kd_{\nu\beta}).$$

As for the terms taken  from  (\ref{R_alfabeta_remain_terms}), they are represented by the first terms on the right of the following equations:  
$$-\tilde d^{\alpha\beta}\hat H_i{\rm \Gamma}^{i}_{\alpha\beta}=\frac12\tilde d^{\alpha\beta}\hat H_i(\tilde h^{im}\tilde{\mathscr D}_m\tilde d_{\alpha\beta})+\frac12\tilde d^{\alpha\beta}\hat H_i(\tilde h^{ia}\tilde{\mathscr D}_a\tilde d_{\alpha\beta}),$$

$$-\tilde d^{\alpha\beta}\Gamma^{k}_{ki}\Gamma^{i}_{\alpha\beta}=\frac12\Gamma^{k}_{ki}(\tilde h^{im}\tilde{\mathscr D}_m\tilde d_{\alpha\beta}+\tilde h^{ia}\tilde{\mathscr D}_a\tilde d_{\alpha\beta}),$$

$$-\tilde d^{\alpha\beta}\Gamma^{a}_{ai}\Gamma^{i}_{\alpha\beta}=\frac12\Gamma^{a}_{ai}(\tilde h^{im}\tilde{\mathscr D}_m\tilde d_{\alpha\beta}+\tilde h^{ia}\tilde{\mathscr D}_a\tilde d_{\alpha\beta}),$$

$$-\tilde d^{\alpha\beta}\Gamma^{\gamma}_{\gamma i}\Gamma^{i}_{\alpha\beta}=\frac12(d^{\gamma\mu}\partial_id_{\gamma\mu})(\tilde h^{im}\tilde{\mathscr D}_m\tilde d_{\alpha\beta}+\tilde h^{ia}\tilde{\mathscr D}_a\tilde d_{\alpha\beta}),$$

$$\tilde d^{\alpha\beta}\Gamma^{\mu}_{\alpha i}\Gamma^{i}_{\mu\beta}=-\frac14(\tilde d^{\gamma\mu}\hat H_i\tilde d_{\gamma\alpha})(\tilde h^{im}\tilde{\mathscr D}_m\tilde d_{\mu\beta}+\tilde h^{ia}\tilde{\mathscr D}_a\tilde d_{\mu\beta}),$$

$$\tilde d^{\alpha\beta}\Gamma^{i}_{\alpha\gamma}\Gamma^{\gamma}_{i\beta}=-\frac14\tilde d^{\alpha\beta}(\tilde h^{im}\tilde{\mathscr D}_m\tilde d_{\alpha\gamma}+\tilde h^{ia}\tilde{\mathscr D}_a\tilde d_{\alpha\gamma})(\tilde d^{\gamma\sigma}\hat H_i\tilde d_{\beta\sigma}).$$
Note that the first term on the right of the previous expression is equal to
$-\frac14\tilde h^{im}d^{\alpha\beta}d^{\gamma\sigma}({\mathscr D}_md_{\alpha\gamma})({\mathscr D}_i d_{\beta\sigma}),$
and 
$$\frac12\tilde d^{\alpha\beta}(\Gamma^{k}_{ki}+\Gamma^{a}_{ai}+\Gamma^{\gamma}_{\gamma i})(\tilde h^{im}\tilde{\mathscr D}_m\tilde d_{\alpha\beta})=\frac12\tilde h^{im}(\Gamma^{k}_{ki}+\Gamma^{a}_{ai}+\Gamma^{\gamma}_{\gamma i})(d^{\alpha\beta}{\partial_m}d_{\alpha\beta}).$$
It can be shown that the first term of $-\tilde d^{\alpha\beta}\hat H_i{\rm \Gamma}^{i}_{\alpha\beta}$ is equal to
$$\frac12(\partial_i\tilde h^{im})(d^{\alpha\beta}\partial_md_{\alpha\beta})+\frac12 \tilde h^{im}\partial_i(d^{\alpha\beta}\partial_md_{\alpha\beta})+\frac12 \tilde h^{im}d^{\alpha\sigma}d^{\mu\beta}({\mathscr D}_id_{\sigma\mu})({\mathscr D}_md_{\alpha\beta}).$$

Using the general relation
\[
 (\partial_{A'}h^{D'E'})=-h^{B'D'}\Gamma ^{E'}_{B'A'}-h^{C'E'}\Gamma ^{D'}_{C'A'}
\]
(here prime indexes mean the following: $A'=(i,a)$),
we can represent $(\partial_i\tilde h^{im})$ in the previous expression as follows:
\[
 \partial_i \tilde h ^{im}=- \tilde h ^{ki}\Gamma^m_{ki}- \tilde h ^{ak}\Gamma^m_{ak}- \tilde h ^{mk}\Gamma^i_{ki}- \tilde h ^{mb}\Gamma^i_{bi}.
\]
Combining the terms of $\tilde R_{ik}\tilde h^{ik}$ and $\tilde R_{\alpha\beta}\tilde d^{\alpha\beta}$ just written out, we get that the first group of elements in $\tilde R$ is defined as
 \begin{eqnarray*}
&&\tilde h^{ik}\partial_i(d^{\alpha\beta}\partial_k d_{\alpha\beta})-\tilde h^{ik}\Gamma^{j}_{ik}(d^{\alpha\gamma}\partial_jd_{\alpha\gamma})-\frac12\tilde h^{ik}\Gamma^{b}_{ik}(d^{\alpha\gamma}\partial_bd_{\alpha\gamma})\\
&&-\frac12\tilde h^{ak}\Gamma^{m}_{ak}(d^{\alpha\gamma}\partial_md_{\alpha\gamma})-\frac12\tilde h^{mb}\Gamma^{i}_{bi}(d^{\alpha\gamma}\partial_md_{\alpha\gamma})+\frac12\tilde h^{im}\Gamma^{a}_{ai}(d^{\alpha\gamma}\partial_md_{\alpha\gamma})\\
&&+\frac14 \tilde h^{im}d^{\alpha\sigma}d^{\mu\beta}({\mathscr D}_id_{\sigma\mu})({\mathscr D}_md_{\alpha\beta})+\frac14\tilde h^{im}(d^{\mu\nu}{\partial_i}d_{\mu\nu})(d^{\alpha\beta}{\partial_m}d_{\alpha\beta}).
\end{eqnarray*}

 The second group of elements in $\tilde R$, the elements of $\tilde R_{ia}\tilde h^{ia}$ and $\tilde R_{\alpha\beta}\tilde d^{\alpha\beta}$,  includes the corresponding parts of 2, 3, 12 terms of (\ref{R_ia_remain_terms}) and  
1, 5, 6, 7, 15, 19 terms of (\ref{R_alfabeta_remain_terms}). Using the same approach as for the elements of the first group, we obtain
\begin{eqnarray*}
&&\tilde h^{ia}\partial_i(d^{\alpha\beta}\partial_a d_{\alpha\beta})-\tilde h^{ia}\Gamma^{b}_{ia}(d^{\alpha\gamma}\partial_bd_{\alpha\gamma})-\frac12\tilde h^{ia}\Gamma^j_{ia}(d^{\alpha\gamma}\partial_jd_{\alpha\gamma})\\
&&-\frac12\tilde h^{mi}\Gamma^{a}_{mi}(d^{\alpha\gamma}\partial_ad_{\alpha\gamma})+\frac12\tilde h^{ia}\Gamma^{b}_{bi}(d^{\alpha\gamma}\partial_ad_{\alpha\gamma})-\frac12\tilde h^{ba}\Gamma^{i}_{bi}(d^{\alpha\gamma}\partial_ad_{\alpha\gamma})\\
&&+\frac14 \tilde h^{im}d^{\alpha\sigma}d^{\mu\beta}({\mathscr D}_id_{\sigma\mu})({\mathscr D}_ad_{\alpha\beta})+\frac14\tilde h^{ia}(d^{\mu\nu}{\partial_i}d_{\mu\nu})(d^{\alpha\beta}{\partial_a}d_{\alpha\beta}).
\end{eqnarray*}

The third group in $\tilde R$ includes the elements of $\tilde R_{ai}\tilde h^{ai}$ and $\tilde R_{\alpha\beta}\tilde d^{\alpha\beta}$. This group can be written as
\begin{eqnarray*}
&&\tilde h^{ai}\partial_a(d^{\alpha\beta}\partial_i d_{\alpha\beta})-\tilde h^{na}\Gamma^{m}_{na}(d^{\alpha\gamma}\partial_md_{\alpha\gamma})-\frac12\tilde h^{ba}\Gamma^m_{ba}(d^{\alpha\gamma}\partial_md_{\alpha\gamma})\\
&&-\frac12\tilde h^{nm}\Gamma^{a}_{na}(d^{\alpha\gamma}\partial_md_{\alpha\gamma})-\frac12\tilde h^{ai}\Gamma^{b}_{ai}(d^{\alpha\gamma}\partial_bd_{\alpha\gamma})+ \frac12\tilde h^{am}\Gamma^{k}_{ka}(d^{\alpha\gamma}\partial_md_{\alpha\gamma})   \\
&&+\frac14 \tilde h^{am}d^{\alpha\sigma}d^{\mu\beta}({\mathscr D}_ad_{\sigma\mu})({\mathscr D}_md_{\alpha\beta})+\frac14\tilde h^{am}(d^{\mu\nu}{\partial_a}d_{\mu\nu})(d^{\alpha\beta}{\partial_m}d_{\alpha\beta}).
\end{eqnarray*}
To get result, we  used the corresponding parts of 2, 3, 12 terms from (\ref{R_ai_remain_terms}) and necessary for us the parts of
2, 8, 9, 10, 18, 20 terms from (\ref{R_alfabeta_remain_terms}).

The fourth group consists of the terms taken from $\tilde R_{ab}\tilde h^{ab}$ and $\tilde R_{\alpha\beta}\tilde d^{\alpha\beta}$.
Proceeding as before, we find that in $\tilde R$, this group of terms is represented as follows:
 \begin{eqnarray*}
&&\tilde h^{ab}\partial_a(\tilde d^{\alpha\beta}\partial_b\tilde d_{\alpha\beta})-\tilde h^{ab}\Gamma^c_{ab}(d^{\alpha\beta}\partial_cd_{\alpha\beta})\\
&&-\frac12 \Bigl(\tilde h^{ia}\Gamma^b_{ia} +\tilde h^{ib}\Gamma^a_{ia}
+\tilde h^{cb}\Gamma^a_{ca}-\tilde h^{ab}\Gamma^k_{ka}-\tilde h^{ab}\Gamma^c_{ca}\Bigr)(d^{\alpha\gamma}\partial_bd_{\alpha\gamma})\\
&&-\frac12 \tilde h^{ab}\Gamma^n_{ab}(d^{\mu\gamma}\partial_nd_{\mu\gamma})\\
&&+ \frac14\tilde h^{ab}d^{\alpha\sigma}d^{\mu\beta}({\mathscr D}_ad_{\sigma\mu})({\mathscr D}_bd_{\alpha\beta})+\frac14\tilde h^{ab}(d^{\alpha\beta}\partial_ad_{\alpha\beta})(d^{\mu\nu}\partial_bd_{\mu\nu}).
\end{eqnarray*}
This group is formed from the corresponding parts of 2, 3, 4, 12  terms from  (\ref{R_ab_remain_terms}) and parts of
2, 8, 9, 10, 18, 20 terms taken from(\ref{R_alfabeta_remain_terms}).

The result of summing all the terms we have received is given by the following expression:
\begin{eqnarray*}
&&\tilde h^{ik}\partial_i(d^{\alpha\beta}\partial_k d_{\alpha\beta})+ \tilde h^{ia}\partial_i(d^{\alpha\beta}\partial_a d_{\alpha\beta})+\tilde h^{ai}\partial_a(d^{\alpha\beta}\partial_i d_{\alpha\beta})+\tilde h^{ab}\partial_a(\tilde d^{\alpha\beta}\partial_b\tilde d_{\alpha\beta})\\
&&-\tilde h^{ik}\Gamma^{j}_{ik}(d^{\alpha\gamma}\partial_jd_{\alpha\gamma})-\tilde h^{ia}\Gamma^{b}_{ia}(d^{\alpha\gamma}\partial_bd_{\alpha\gamma})-\tilde h^{na}\Gamma^{m}_{na}(d^{\alpha\gamma}\partial_md_{\alpha\gamma})\\
&&-\tilde h^{ab}\Gamma^c_{ab}(d^{\alpha\beta}\partial_cd_{\alpha\beta})-(\tilde h^{ik}\Gamma^{b}_{ik}\partial_b+\tilde h^{ak}\Gamma^{m}_{ak}\partial_m+\tilde h^{ai}\Gamma^{b}_{ai}\partial_b+\tilde h^{ab}\Gamma^{m}_{ab}\partial_m)\ln d\\
&&+\frac14\tilde h^{ik}(\partial_i\ln d)(\partial_k \ln d)+\frac14\tilde h^{ia}(\partial_i\ln d)(\partial_a \ln d)\\
&&+\frac14\tilde h^{am}(\partial_a\ln d)(\partial_m \ln d)+\frac14\tilde h^{ab}(\partial_a\ln d)(\partial_b \ln d)\\
&&+\frac14\tilde h^{im}d^{\alpha\sigma}d^{\mu\beta}({\mathscr D}_id_{\sigma\mu})({\mathscr D}_md_{\alpha\beta})+\frac14 \tilde h^{im}d^{\alpha\sigma}d^{\mu\beta}({\mathscr D}_id_{\sigma\mu})({\mathscr D}_ad_{\alpha\beta})\\
&&+\frac14 \tilde h^{am}d^{\alpha\sigma}d^{\mu\beta}({\mathscr D}_ad_{\sigma\mu})({\mathscr D}_md_{\alpha\beta})+ \frac14\tilde h^{ab}d^{\alpha\sigma}d^{\mu\beta}({\mathscr D}_ad_{\sigma\mu})({\mathscr D}_bd_{\alpha\beta}),
\end{eqnarray*}
where $d=\det d_{\alpha\beta}$. The obtained expression can be rewritten as
\begin{eqnarray}
&&\bigl(\tilde h^{ik}\partial_i\partial_k + \tilde h^{ia}\partial_i\partial_a +\tilde h^{ai}\partial_a\partial_i +\tilde h^{ab}\partial_a\partial_b)\bigr.\nonumber\\
&&-\tilde h^{ik}\Gamma^{j}_{ik}\partial_j-\tilde h^{ia}\Gamma^{b}_{ia}\partial_b-\tilde h^{na}\Gamma^{m}_{na}\partial_m-\tilde h^{ab}\Gamma^c_{ab}\partial_c\nonumber\\
&&-\bigl.\tilde h^{ik}\Gamma^{b}_{ik}\partial_b-\tilde h^{ak}\Gamma^{m}_{ak}\partial_m-\tilde h^{ai}\Gamma^{b}_{ai}\partial_b-\tilde h^{ab}\Gamma^{m}_{ab}\partial_m\bigr)\ln d\nonumber\\
&&+\frac14\tilde h^{ik}(\partial_i\ln d)(\partial_k \ln d)+\frac14\tilde h^{ia}(\partial_i\ln d)(\partial_a \ln d)\nonumber\\
&&+\frac14\tilde h^{am}(\partial_a\ln d)(\partial_m \ln d)+\frac14\tilde h^{ab}(\partial_a\ln d)(\partial_b \ln d)\nonumber\\
&&+\frac14\tilde h^{im}d^{\alpha\sigma}d^{\mu\beta}({\mathscr D}_id_{\sigma\mu})({\mathscr D}_md_{\alpha\beta})+\frac14 \tilde h^{im}d^{\alpha\sigma}d^{\mu\beta}({\mathscr D}_id_{\sigma\mu})({\mathscr D}_ad_{\alpha\beta})\nonumber\\
&&+\frac14 \tilde h^{am}d^{\alpha\sigma}d^{\mu\beta}({\mathscr D}_ad_{\sigma\mu})({\mathscr D}_md_{\alpha\beta})+ \frac14\tilde h^{ab}d^{\alpha\sigma}d^{\mu\beta}({\mathscr D}_ad_{\sigma\mu})({\mathscr D}_bd_{\alpha\beta}).
\label{fourgroup}
\end{eqnarray}
The first terms of this expression can be presented in the following form:
$$\triangle_{\scriptscriptstyle \tilde {\mathcal M}}\ln d+\frac14 G_{\scriptscriptstyle \tilde{\mathcal M}}^{A'B'}(\partial_{A'}\ln d)(\partial_{B'}\ln d),$$
where $\triangle_{\scriptscriptstyle \tilde {\mathcal M}}$ is the Laplace-Beltrami operator on $\tilde {\mathcal M}$:
$$\triangle_{\scriptscriptstyle \tilde {\mathcal M}}=\tilde h^{A'B'}\partial_{A'}\partial_{B'}-\tilde h^{A'B'}\Gamma^{C'}_{A'B'}\partial_{C'},$$
and the  prime indexes also  mean the following:  $A'=(i,a)$.

Taking into account the fact that, in addition to the terms (\ref{fourgroup}) we have obtained, the scalar curvature ${\rm R}_{\tilde{\mathcal M}}$, the scalar curvature of the orbit ${\rm R}_{\mathcal G}$, and  the terms with the curvature $\mathscr F$ also contribute to the scalar curvature ${\rm R}_{\tilde{\mathcal P}}$, we will have
\begin{eqnarray}
{\rm R}_{\tilde{\mathcal P}}&=&{\rm R}_{\tilde{\mathcal M}}+{\rm R}_{\mathcal G}+\frac14 \tilde h^{A'B'}\tilde h^{C'D'}d_{\mu\nu}{\mathscr F}^{\mu}_{A'C'}{\mathscr F}^{\nu}_{B'D'}\nonumber\\
&&+\frac14\tilde h^{A'B'}d^{\mu\sigma}d^{\nu\kappa}({\mathscr D}_{A'}d_{\mu\nu})({\mathscr D}_{B'}d_{\sigma\kappa})+\triangle_{\scriptscriptstyle \tilde {\mathcal M}}\ln d+\frac14G_{\scriptscriptstyle \tilde{\mathcal M}}(\partial \ln d,\partial \ln d).
\nonumber\\
\label{curvat_itog}
\end{eqnarray}

\appendix
\section*{Appendix C}
\section*{The second quadratic form of the orbit}
\setcounter{equation}{0}
\def\theequation{C.\arabic{equation}}

\subsection*{Identities}
For the initial metric $ds^2=G_{AB}(Q)dQ^AdQ^B+G_{ab}df^Adf^b$ and the metric on the orbit $d_{\alpha\beta}(Q,f)=K^A(Q)G_{AB}(Q)K^B_{\beta}(Q)+K^a(f)G_{ab}K^b_{\beta}(f),$
the following identities can be easily obtained:
\begin{eqnarray*}
-G^{EC}(Q)\frac{\partial d_{\alpha\beta}(Q,f)}{\partial Q^C}&=&\Bigl[({\nabla}_{K_{\alpha}}K_{\beta})^E(Q)+({\nabla}_{K_{\beta}}K_{\alpha})^E(Q)\Bigr],
\nonumber\\
-G^{ab}\frac{\partial d_{\alpha\beta}(Q,f)}{\partial f^b}&=&\Bigl[({\nabla}_{K_{\alpha}}K_{\beta})^a(f)+({\nabla}_{K_{\beta}}K_{\alpha})^a(f)\Bigr].
\end{eqnarray*}

In the variables $(x^i,\tilde f^b,a^{\alpha})$,  these identities looks as follows:
\begin{eqnarray}
&&\frac12G^{CE}(Q^{\ast}(x))\Bigl[ G^{\rm H}_{CD}Q^{\ast D}_mh^{mi}\frac{\partial d_{\alpha\beta}(Q^{\ast}(x),\tilde f)}{\partial x^i}-\Lambda^{\beta}_CK^a_{\beta}(\tilde f)\frac{\partial d_{\alpha\beta}(Q^{\ast}(x),\tilde f)}{\partial \tilde f^a}
\nonumber\\
&&+\Lambda ^{\varepsilon}_C(c^{\varphi}_{\varepsilon \mu}d_{\varphi \nu}+c^{\varphi}_{\varepsilon \nu}d_{\varphi \mu})\Bigr]
=-\frac12\Bigl[({\nabla}_{K_{\alpha}}K_{\beta})^E(Q^{\ast}(x))+({\nabla}_{K_{\beta}}K_{\alpha})^E(Q^{\ast}(x))\Bigr],
\nonumber\\
\label{ident_Q_ast}
\end{eqnarray}
\begin{equation}
\frac12G^{pq}\frac{\partial}{\partial \tilde f^q}d_{\alpha\beta}(Q^{\ast}(x),\tilde f)=-\frac12\Bigl[({\nabla}_{K_{\alpha}}K_{\beta})^p(\tilde f)+({\nabla}_{K_{\beta}}K_{\alpha})^p(\tilde f)\Bigr].
\label{ident_f_tild}
\end{equation} 
$$(d_{\alpha\beta}(Q,f)=\rho^{\alpha'}_{\alpha}\rho^{\beta'}_{\beta}d_{\alpha'\beta'}(Q^{\ast}(x),\tilde f)=\rho^{\alpha'}_{\alpha}\rho^{\beta'}_{\beta}(\gamma_{\alpha'\beta'}(Q^{\ast}(x))+\gamma'_{\alpha'\beta'}(\tilde f)).)$$
Note that in these identities, the expressions in square brackets to the right were denoted  in the main text as $(\dots)^E$ and  $(\dots)^p.$

\subsection*{Calculation of $j(1)_{\alpha\beta}$}
$$j(1)_{\alpha\beta}=\Bigl[\tilde h^{kn}\tilde G^{\rm H}_{BM}Q^{\ast B}_k(\dots)^M+\tilde h^{kn}\tilde G^{\rm H}_{Ab}Q^{\ast A}_k(\dots)^b\Bigr]\hat{H}_n.$$
Using the expression  for $(\dots)^M$  from the  indentity (\ref{ident_Q_ast}) in the first term of $j(1)_{\alpha\beta}$, we rewrite this term as follows:
$$-\frac12\tilde h^{kn}G^{MC}\tilde G^{\rm H}_{BM}Q^{\ast B}_k\Bigl[ G^{\rm H}_{CD}Q^{\ast D}_mh^{mi}\frac{\partial d_{\alpha\beta}}{\partial x^i}-\Lambda^{\beta}_CK^a_{\beta}(\tilde f)\frac{\partial d_{\alpha\beta}}{\partial \tilde f^a}
+\Lambda ^{\varepsilon}_C(c^{\varphi}_{\varepsilon \mu}d_{\varphi \nu}+c^{\varphi}_{\varepsilon \nu}d_{\varphi \mu})\!\Bigr].$$

Since $G^{MC}G^{\rm H}_{CD}={\Pi}^M_D$, $\tilde G^{\rm H}_{BM}{\Pi}^M_D= G^{\rm H}_{BD}$ and $Q^{\ast B}_kG^{\rm H}_{BD}Q^{\ast D}_m=h_{km}$, the first term of the previous expression  is equal to $$-\frac12\tilde h^{jn}\frac{\partial}{\partial x^j}d_{\alpha\beta}.$$

In the second term, 
$-\frac12\tilde h^{kn}G^{MC}\tilde G^{\rm H}_{BM}Q^{\ast B}_k(-\Lambda^{\beta}_C)K^a_{\beta}(\tilde f)\frac{\partial}{\partial \tilde f^a}d_{\alpha\beta}$, we have
$\;\;G^{MC}\tilde G^{\rm H}_{BM}=\tilde{\Pi}^C_B$ and $\tilde{\Pi}^C_BQ^{\ast B}_k\Lambda^{\beta}_C=\mathscr A^{\beta}_k$. So, this term is equal to $$-\frac12 \tilde h^{kn}\mathscr A^{\beta}_kK^a_{\beta}\frac{\partial}{\partial \tilde f^a}d_{\alpha\beta}.$$
Similarly, for the last term we get 
$\frac12 \tilde h^{kn}\mathscr A^{\varepsilon}_k(c^{\varphi}_{\varepsilon \mu}d_{\varphi \nu}+c^{\varphi}_{\varepsilon \nu}d_{\varphi \mu}).$

Using the identity (\ref{ident_f_tild}) for $(\dots)^b$ in $j(1)_{\alpha\beta}$, we get that
 this term is equal to
$$\frac12\tilde h^{kn}\mathscr A^{\nu}_kK^c_{\nu}\frac{\partial}{\partial \tilde f^c}d_{\alpha\beta}.$$
(Note that
 $-\frac12\tilde h^{kn}Q^{\ast A}_k(-G_{AR}K^R_{\mu}d^{\mu\nu})K^c_{\nu}=\frac12\tilde h^{kn}\mathscr A^{\nu}_kK^c_{\nu}.$) 
 
 Combining the obtained expressions we come to the following representation for $j(1)_{\alpha\beta}$:
$$j(1)_{\alpha\beta}=-\frac12\tilde h^{kn}(\mathscr D_kd_{\alpha\beta})\hat H_n.$$

\subsection*{Calculation of $j(2)_{\alpha\beta}$}

$$j(2)_{\alpha\beta}=\tilde h^{kb}\Bigl[\Bigl( G^{\rm H}_{ML}Q^{\ast L}_mh^{mi}\tilde h_{ki}+N^a_M\tilde h_{ak}\Bigr)(\dots)^M+\tilde h_{ck}(\dots)^c\Bigr]\hat{H}_b,$$
where for the metric on 
$\tilde{\mathcal M}$ we have
$$\tilde h^{kb}=G^{EF}N^b_FN^P_ET^k_P,\;\;\;\;\tilde h_{ki}=Q^{\ast A}_k\tilde G^{\rm H}_{AB}Q^{\ast B}_i,\;\;\;\; \tilde h_{ak}=\tilde G^{\rm H}_{Ba}Q^{\ast B}_k,\;\;\;\;\;h_{ab}=\tilde G^{\rm H}_{ab}.$$
 The operator $T^i_A$ is defined as 
\[
T^i_A=(P_{\bot})^D_A(Q^{\ast}(x))G^H_{DL}(Q^{\ast}(x))Q^{\ast L}_m(x)h^{mi}(x).
\] 
It has two important properties: $T^i_A Q^{\ast A}_k={\delta}^i_k$ and $Q^{\ast A}_jT^j_B=(P_{\bot})^A_B$. 

First, we transform  terms placed in parentheses which followed by $(\dots)^M$:
\begin{eqnarray*}
G^{\rm H}_{ML}Q^{\ast L}_mh^{mi}Q^{\ast A}_k\tilde G^{\rm H}_{AB}Q^{\ast B}_i+N^a_M\tilde G^{\rm H}_{Ba}Q^{\ast B}_k&=&
\nonumber\\
=(N^B_M\tilde G^{\rm H}_{AB}+N^a_M\tilde G^{\rm H}_{Aa})Q^{\ast A}_k&=&\tilde G^{\rm H}_{MA}Q^{\ast A}_k.
\end{eqnarray*}
Then $j(2)$ is rewritten as
$$\tilde h^{kb}\Bigl[\tilde G^{\rm H}_{AM}Q^{\ast A}_k(\dots)^M+\tilde G^{\rm H}_{Bc}Q^{\ast B}_k(\dots)^c\Bigr]\hat H_b.$$
Using the identities  (\ref{ident_Q_ast}) and (\ref{ident_f_tild}) for $(\dots)^M$ and $(\dots)^c$ in this expression, and then transforming the result, just like for $j(1)$, we  obtain
$$j(2)_{\alpha\beta}=-\frac12\tilde h^{kb}(\mathscr D_kd_{\alpha\beta})\hat H_b.$$

\subsection*{Calculation of $j(3)_{\alpha\beta}$}
$$j(3)_{\alpha\beta}=\tilde h^{bk}\Bigl[\Bigl( G^{\rm H}_{ML}Q^{\ast L}_mh^{mi}\tilde h_{ib}+N^a_M\tilde h_{ab}\Bigr)(\dots)^M+\tilde h_{cb}(\dots)^c\Bigr]\hat{H}_k.$$

To calculate $j(3)_{\alpha\beta}$, we proceed in the same way as in the case    of $j(1)_{\alpha\beta}$ and $j(2)_{\alpha\beta}$. First, we also transform  the terms placed in parenthesis.
 But now  we use the following  property:
$N^B_M\tilde G^{\rm H}_{Bb}+N^a_M\tilde G^{\rm H}_{ab}=\tilde G^{\rm H}_{Mb}$. Therefore, in this case
$$j(3)_{\alpha\beta}=\tilde h^{bk}\Bigl(\tilde G^{\rm H}_{Mb}(\dots)^M+\tilde G^{\rm H}_{db}(\dots)^d\Bigr)\hat H_k.$$

Next, we use the identity  (\ref{ident_Q_ast}) for $(\dots)^M$. After substitution, the first term of $j(3)_{\alpha\beta}$ looks as follows:
$$-\frac12\tilde h^{bk}G^{MC}\tilde G^{\rm H}_{Mb}\Bigl[ G^{\rm H}_{CD}Q^{\ast D}_mh^{mi}\frac{\partial}{\partial x^i}d_{\alpha\beta}-\Lambda^{\beta}_CK^a_{\beta}(\tilde f)\frac{\partial}{\partial \tilde f^a}d_{\alpha\beta}
+\Lambda ^{\varepsilon}_C(c^{\varphi}_{\varepsilon \mu}d_{\varphi \nu}+c^{\varphi}_{\varepsilon \nu}d_{\varphi \mu})\Bigr]$$

The first term of this expression vanishes, since $G^{MC} G^{\rm H}_{CD}={\Pi}^M_D$  and $\tilde G^{\rm H}_{Mb}{\Pi}^M_D=0$.
The second term can be presented as $\;\;-\frac12\tilde h^{bk}{\mathscr A}^{\beta}_bK^a_{\beta}
\frac{\partial}{\partial \tilde f^a}d_{\alpha\beta}$.
The third term is equal to $\;\;\frac 12\tilde h^{bk}{\mathscr A}^{\varepsilon}_b(c^{\varphi}_{\varepsilon \mu}d_{\varphi \nu}+c^{\varphi}_{\varepsilon \nu}d_{\varphi \mu}).$

Using the identity (\ref{ident_f_tild}) for $(\dots)^d$  in $\tilde h^{bk}\tilde G^{\rm H}_{db}(\dots)^d(\tilde f)$, the second term of $j(3)_{\alpha\beta}$, we get
$$-\frac12\tilde h^{bk}\tilde G^{\rm H}_{db}G^{da}\frac{\partial\, d_{\alpha\beta}}{\partial \tilde f^a}=-\frac12\tilde h^{bk}\tilde{\Pi}^a_b\frac{\partial\, d_{\alpha\beta}}{\partial \tilde f^a}=-\frac12\tilde h^{bk}\frac{\partial\, d_{\alpha\beta}}{\partial \tilde f^a}+\frac12\tilde h^{bk}{\mathscr A}^{\mu}_bK^a_{\mu}\frac{\partial\, d_{\alpha\beta}}{\partial \tilde f^a}.$$
Combining the previous expressions, we get that $j(3)_{\alpha\beta}$ can be represented as follows:
$$j(3)_{\alpha\beta}=-\frac12\tilde h^{bk}(\mathscr D_b d_{\alpha\beta})\hat H_k.$$

\subsection*{Calculation of $j(4)_{\alpha\beta}$}
$$j(4)_{\alpha\beta}={\tilde h}^{ab}\Bigl[\tilde G^{\rm H}_{Ma}(\dots)^M+\tilde G^{\rm H}_{da}(\dots)^d\Bigr]\hat{H}_b.$$

Replacing $(\dots)^M$  in $j(4)_{\alpha\beta}$ with the help of the identity  (\ref{ident_Q_ast}), we obtain, as previously, three terms. 
 It can be shown that the term with the derivative $\frac{\partial \;d_{\alpha\beta}}{\partial x^i}$ vanishes. This follows from the equality  $\tilde G^{\rm H}_{Ma}{\Pi}^M_D=0$. 
The second term is equal to  $-\frac12{\tilde h}^{ab}{\mathscr A}^{\beta}_aK^d_{\beta}\,(\partial d_{\alpha\beta}/{\partial \tilde f^d})$.
The third term of this replacement is $\frac12{\tilde h}^{ab}{\mathscr A}^{\varepsilon}_a(c^{\varphi}_{\varepsilon \mu}d_{\varphi \nu}+c^{\varphi}_{\varepsilon \nu}d_{\varphi \mu})$.

On the other hand,  using the identity  (\ref{ident_f_tild}) for $(\dots)^d$ in $j(4)_{\alpha\beta}$ , we find that  ${\tilde h}^{ab}\tilde G^{\rm H}_{da}(\dots)^d=-\frac12 {\tilde h}^{ab}\Bigl(\frac{\partial \;d_{\alpha\beta}}{\partial \tilde f^a}-{\mathscr A}^{\mu}_aK^p_{\mu}\frac{\partial \;d_{\alpha\beta}}{\partial \tilde f^p}\Bigr).$ As a result, we get
$$j(4)_{\alpha\beta}=-\frac12\tilde h^{ab}(\mathscr D_a d_{\alpha\beta})\hat H_b.$$

\end{document}